\documentclass[12pt,leqno]{article}
\usepackage[T1]{fontenc}
\usepackage{graphicx}
\usepackage{caption}
\usepackage{subcaption}
\usepackage{mathrsfs}
\usepackage{amsthm}
\usepackage{epsfig}
\usepackage{latexsym}
\usepackage{amssymb}
\usepackage{amsfonts}
\usepackage{amsmath}
\usepackage{euscript}
\usepackage{hyperref}
\hypersetup{
   colorlinks,
    citecolor=blue,
    filecolor=black,
    linkcolor=red,
    urlcolor=magenta
}
\hypersetup{linktocpage}
\newcommand{\C}{\mathbb{C}}

\newcommand{\R}{\mathbb{R}}
\newcommand{\N}{\mathbb{N}}

\renewcommand{\d}{\mathrm{d}}

\newcommand{\dvol}{\mathrm{dVol}}
\newcommand{\Log}{\mathrm{Log}}
\newcommand{\doc}{{\cal D}_{oc}}

\newtheorem{definition}{Definition}[section]

\newtheorem{lemma}{Lemma}[section]
\newtheorem{remark}{Remark}[section]

\begin{document}\mbox{}

\vspace{0.25in}

\begin{center}
{\bf {\huge Superradiance on the Reissner-Nordstr\o m metric}}

\vspace{0.25in}

{\large Laurent DI MENZA\footnote{Laboratoire de Math\'ematiques de Reims, FR CNRS 3399, UFR SEN, Moulin de la Housse, BP 1039, 51687 Reims Cedex 2, France.\\
E-mail: Laurent.Di-Menza@univ-reims.fr} and Jean-Philippe NICOLAS\footnote{LMBA, UMR CNRS n$^\mathit{o}$ 6205, Université de Brest, 6 avenue Le Gorgeu, 29238 Brest cedex 3, France. \\
E-mail: Jean-Philippe.Nicolas@univ-brest.fr}}
\end{center}

\vspace{0.1in}

\begin{center}
{\bf Abstract}
\end{center}

In this article, we study the superradiance of charged scalar fields on the sub-extremal Reissner-Nordstr\o m metric, a mechanism by which such fields can extract energy from a static spherically symmetric charged black hole. A geometrical way of measuring the amount of energy extracted is proposed. Then we investigate the question numerically. The toy-model and the numerical methods used in our simulations are presented and the problem of long time measurement of the outgoing energy flux is discussed. We provide a numerical example of a field exhibiting a behaviour analogous to the Penrose process~: an incoming wave packet which splits, as it approaches the black hole, into an incoming part with negative energy and an outgoing part with more energy than the initial incoming one. We also show another type of superradiant solution for which the energy extraction is more important. Hyperradiant behaviour is not observed, which is an indication that the Reissner-Nordstr\o m metric is linearly stable in the sub-extremal case.


\section{Introduction}

Although the existence of black holes was already conjectured in the {\sc XVII}I$^\mathrm{th}$ century by Mitchell and Laplace, it is in the {\sc XX}$^\mathrm{th}$ century, within the framework of general relativity, that these objects finally acquired an unambiguous mathematical status, first as explicit solutions of the Einstein equations (the Schwarzschild solution in 1917, the Kerr solution in 1963) then as inevitable consequences of the evolution of the universe (assuming matter satisfies the dominant energy condition and a sufficient amount of it is concentrated in a small enough domain, see S. Hawking and R. Penrose, 1970 \cite{HaPe}). It is in the neighbourhood of black holes that general relativity reveals its most striking aspects. One of the remarkable phenomena occurring in such regions is superradiance~: the possibility for bosonic fields to extract energy from the black hole. This energy extraction can result from the interaction of the charges of the black hole and the field, or from the tidal effects of rotation. Superradiance has very different features in each of these two cases. For rotation induced superradiance, the energy extraction is localized in a fixed neighbourhood of the horizon called the ergosphere. In the case of charge interaction, the region where energy extraction takes place varies with the physical parameters of the field (mass, charge, angular momentum) and may even cover the whole exterior of the black hole. We shall refer to it as the ``effective ergosphere'', following G. Denardo and R. Ruffini \cite{DeRu} who first introduced this notion\footnote{The effective, or generalized, ergosphere should not be confused with the dyadosphere, introduced by G. Preparata, R. Ruffini and Xue S.-S. \cite{PreRuXu}, which is the region outside the horizon of an electromagnetic black hole where the electromagnetic field exceeds the critical value for electron-positron pair
production.}. A question of crucial importance concerning the stability of black hole spacetimes is the amount of energy that can be extracted. If the process turned out not to be limited and fields could extract an infinite amount of energy, this would indicate that when taking the back reaction of the field on the metric into account, the evolution could be unstable.

In the case of rotationally induced superradiance, although the problems are very delicate, the clear-cut geometrical nature of the ergosphere allows the use of geometric-analytic methods such as vector field techniques in order to study the evolution of fields. Moreover, in the case of Kerr black holes, the separability of the equations provides a convenient access to the frequency analysis of their solutions. Hence, superradiance outside a Kerr black hole has been directly or indirectly studied in various papers. The pioneering work was done by S.A. Teukolsky and W.H. Press in \cite{PreTe2}, the third in a series of papers analyzing the perturbations of a Kerr black hole by spin $1/2$, $1$ and $2$ fields (\cite{Teu,PreTe1,PreTe2} in chronological order), using the Newman-Penrose formalism and the separability of the equations~; it contained a precise description and numerical analysis of superradiant scattering processes. In 1983, S. Chandrasekhar's remarkable book \cite{Cha} provided a complete and up-to-date account of the theory of black hole perturbations, with in particular calculations of scattering matrices (reflection and transmission coefficients) at fixed frequencies for Reissner-Nordstr\o m and Kerr black holes. The last decade or so has seen a rather intensive analytical study of superradiance around rotating black holes (see for example \cite{AB, DR, FKSY, Ha}). It is now clear that in the slowly rotating case, the amount of energy extracted by scalar fields is finite and controlled by a fixed multiple of the initial energy of the field. Recent results \cite{AB2} show that the phenomenon is similar for Maxwell fields. In a different spirit, conditions on the stress-energy tensor under which fields or matter outside a Kerr black hole can extract rotational energy are derived in a recent paper by J.-P. Lasota, E. Gourgoulhon, M. Abramowicz, A. Tchekhovskoy, R. Narayan \cite{EGou}~; these extend the conditions for energy extraction in the Penrose process.

In the case of charge-induced superradiance, we lose the covariance of the equations and the niceties of a strong geometrical structure. To our knowledge, the only example of a theoretical approach to the question is a work of A. Bachelot \cite{Ba} where a general spherically symmetric situation is considered, a particular case of which is charged scalar fields outside a De Sitter-Reissner-Nordstr\o m black hole. A rigorous definition of two types of energy extracting modes is provided~: superradiant modes, where the energy extracted is finite, and hyperradiant modes corresponding to the extraction of an infinite amount of energy. We give an excerpt from \cite{Ba} which is interesting for the clear definition of the two types of modes but also for the last sentence as we shall see below~: {\it ``when $\kappa$ is a superradiant mode, $\vert R^\pm (\kappa)\vert$ is strictly larger than $1$, but finite, this is the phenomenon of superradiance of the Klein-Gordon fields. [...] $T^\pm$ and $R^\pm$ diverge at the hyperradiant modes\footnote{$T^\pm$ and $R^\pm$ are the transmission and the reflection coefficients for a full scattering experiment, i.e. both incoming and outgoing conditions are considered at the horizon.}. The situation differs for the Dirac or Schrödinger equations, for which the reflection is total in the Klein zone.''} A. Bachelot's work also provides a criterion characterizing the presence of hyperradiant modes. Unfortunately, it involves the existence of zeros of a particular analytic function and in the De Sitter-Reissner-Nordstr\o m case seems difficult to apply due to the complexity of the function to study.

When it comes to numerical investigation of superradiance by charged black holes, very few results are available in the literature. M. Richartz and A. Saa \cite{RiSa} work in the frequency domain, i.e. via a Fourier analysis in time. The scattering process is then described at a given frequency by the scattering matrix, whose coefficients are the reflection and transmission coefficients. Their paper provides an expression of the transmission coefficient for spin $0$ and spin $1/2$ fields outside a Reissner-Nordstr\o m black hole, in the small frequency regime. Very recently, a thorough review on superradiance has appeared, by R. Brito, V. Cardoso and P. Pani \cite{BriCaPa}, with in particular a frequency domain analysis of the superradiance of charged scalar fields by a Reissner-Nordstr\o m black hole. Another interesting approach is to study superradiance in the time domain, i.e. to follow the evolution of the field without performing a Fourier transform in time. With this approach, the natural thing to look for is a field analogue of the Penrose process. Let us recall here that the Penrose process is a thought experiment proposed by R. Penrose \cite{Pe1969} then discussed in more details in R. Penrose and R.M. Floyd \cite{PeFlo}, by which a particle is sent towards a rotating black hole, enters the ergosphere, disintegrates there into a particle with negative energy, which falls into the black hole, and another particle, which leaves the ergosphere again with more energy than the incoming one. A field analogue of this mechanism would involve wave packets instead of particles, the rest being essentially unchanged. In particular the incoming wave packet needs, when entering the ergosphere, to split spontaneously into a wave packet with negative energy (which is then bound to fall into the black hole or at least stay in the region where negative energy is allowed) and another one which propagates to infinity with more energy than the initial incoming one. To this day and to our knowledge, charge-induced superradiance has not been studied in the time-domain, but there are some important recent results that use this approach in the Kerr case, due to P. Czismadia, A. László and I. Rácz \cite{CziLaRa} and to A. L\'aszl\'o and I. R\'acz \cite{LaRa}. What the authors observe, in the case of the wave equation outside a rather fast rotating Kerr black hole, is that the phenomenon described above does not seem to occur~; instead their simulations show that wave packets that are a priori in the superradiant regime, undergo an almost complete reflection as they enter the ergosphere. This is in sharp contrast with the excerpt from \cite{Ba} in the previous paragraph. But of course one must be careful when comparing things that are not directly comparable~; A. Bachelot in \cite{Ba} focuses on a general spherically symmetric setting, while in \cite{CziLaRa,LaRa} P. Czismadia, A. L\'aszl\'o and I. R\'acz work in the case of a rotating black hole which is not even slowly rotating and cannot therefore be seen as a small perturbation of a spherically symmetric situation.

The present paper focuses on a case of charge-induced superradiance~: the evolution of charged scalar fields outside a sub-extremal Reissner-Nordstr\o m black hole, which we study in the time domain. It is organized as follows. Section \ref{AnDeToMo} contains a description of superradiance purely as a problem of analysis of partial differential equations and a reduction to a toy model. The natural conserved energy is not positive definite and the region where the energy density is allowed to become negative depends on the choice of physical parameters. In section \ref{EnergyCurrent}, after a short account of Reissner-Nordstr\o m metrics and the definition of the charged D'Alembertian on such backgrounds, we construct a conserved energy current. The field equation that we study is not covariant because the electromagnetic field is not influenced in return by the scalar field, so there is no way of constructing a conserved stress-energy tensor. However, we introduce a simple modification of the stress-energy tensor for a Klein-Gordon field, which, though not conserved, leads to a conserved current when we contract it with the global timelike Killing vector field. This gives us a geometrical means of measuring the amount of energy extracted by a given field. Section \ref{Numerics} contains the numerical investigation. The algorithm is described, with a particular attention paid to the treatment of the boundary. This is followed by a detailed study of the toy model, with the hyperradiant and superradiant cases. Then the full problem is tackled. Our observations indicate the absence of hyperradiant modes and the existence of superradiant modes. We exhibit two types of superradiant solutions. The first kind is the field analogue of the Penrose process~: incoming wave packets splitting in the correct manner and coming out of the effective ergosphere with more energy than they brought in. Our results show an energy extraction of nearly $50\%$ (i.e. a gain, or reflection coefficient, close to $1.5$). We do not observe the total reflection seen in \cite{CziLaRa,LaRa} and our simulations are consistant with the general spherically symmetric analysis in \cite{Ba} and the numerical simulations in \cite{BriCaPa}. The second kind is given by solutions whose data are located within the effective ergosphere but on which the energy is nevertheless positive definite~; we call them flare-type initial data because the field vanishes at $t=0$ but not its time derivative. This second kind of solution provides much larger energy gains. For both these classes of solutions, the gain always stabilizes to a finite value, ruling out hyperradiant behaviour. As a test of the robustness of our numerical scheme, we push it into the high energy regime~; we observe exactly as expected the concentration of the propagation of the field along the null geodesics as well as the fact that superradiance is a low energy phenomenon. We discuss our results in the conclusion.

\section{Analytic description and toy model} \label{AnDeToMo}

The evolution of a charged scalar field $\phi$ with charge $q$, mass $m$ and angular momentum $l\in \N$ (meaning that $-\Delta_{S^2} \phi = l(l+1) \phi$) outside a Reissner-Nordstr\o m black hole is described by the equation on $\R_t \times \R_x$
\begin{equation} \label{CKGCoord}
\left\{ \frac{\partial^2}{\partial t^2} -2i\frac{qQ}{r}
  \frac{\partial}{\partial t}
  - \frac{\partial^2}{\partial x^2} + F(r) \left(
    \frac{l(l+1)}{r^2} + m^2 + \frac{F'(r)}{r} \right) - \frac{q^2
  Q^2}{r^2} \right\} \phi   =0 \, ,
\end{equation}
where $Q\neq 0$ and $M> \vert Q \vert$, are the charge and mass of the black hole, $r$ is determined in terms of $x$ by $r = G(x)$ where $G$ is an analytic function\footnote{It is the reciprocal function of $r\mapsto r_*$ defined in \eqref{RstarNonExt} and \eqref{RstarExt} in the sub-extremal and extreme cases.} on $\R$, strictly increasing, such that
\[ G(x) \simeq x \mbox{ as } x \rightarrow +\infty \, ,~ G(x) \rightarrow r_+ = M + \sqrt{M^2 -Q^2} >0 \mbox{ as } x \rightarrow -\infty \]
and the function $F(r)$ is given by
\[ F(r) = 1 - \frac{2M}{r} + \frac{Q^2}{r^2} \, .\]
The region $x \rightarrow -\infty$ corresponds to the neighbourhood of the event horizon localized at $r=r_+$~; the function $F$ is positive on $] r_+ , +\infty [$ and vanishes at $r_+$.

This equation has a natural conserved quantity given by
\begin{equation} \label{ConservedEnergy}
 \frac12 \int_{\R} \left\{ \left| \partial_t \phi \right|^2 + \left|
  \partial_{x} \phi \right|^2 + \left( F(r) \left(
    \frac{l(l+1)}{r^2} + m^2 + \frac{F'(r)}{r} \right) - \frac{q^2
  Q^2}{r^2} \right)
  \left| \phi \right|^2 \right\} \d x \, .
  \end{equation}
Since $F$ vanishes at $r_+$, unless the field is uncharged the potential
\begin{equation} \label{PotTot}
F(r) \left( \frac{l(l+1)}{r^2} + m^2 + \frac{F'(r)}{r} \right) - \frac{q^2 Q^2}{r^2}
\end{equation}
becomes negative near the horizon and the conserved quantity is therefore not positive definite.
\begin{definition}
The effective ergosphere is the region where the potential \eqref{PotTot} becomes negative.
\end{definition}
The localization of the effective ergosphere depends on the respective values of $q$, $l$, $m$, $Q$ and $M$ in a complicated way. In the worst cases, it covers a neighbourhood of infinity, maybe even the whole real axis. As soon as the mass $m$ of the field is positive, the potential is positive in a neighbourhood of infinity and the effective ergosphere is localized in a compact region around the black hole. This is also the case when the mass of the field vanishes and its angular momentum is large enough, namely $l(l+1) > q^2 Q^2$. In the case where $m=0$ and $l(l+1) \leq q^2 Q^2$ (the case $q=0$ is not interesting for us since there is no superradiance), the effective ergosphere covers a neighbourhood of infinity.

Note that the mass of the field, when it is large enough, prevents the occurrence of super-radiance in the sense that there exists a positive definite conserved energy. This is well-known and easy to prove but does not seem to be present in the literature, apart from a remark in \cite{Ba} p. 1207. We state and prove the result here.
\begin{lemma}
For $m\geq \left| \frac{qQ}{r_+} \right|$, there is no superradiance in
the sense that there exists another conserved energy that is
positive definite outside the black hole.
\end{lemma}
\noindent{\bf Proof.} Putting
\[ \psi:= e^{-i\frac{qQ}{r_+}t}\phi \, ,\]
we find that $\psi$ satisfies the equation
\[ \left\{ \left( \frac{\partial}{\partial t} -iqQ \left( \frac{1}{r}
- \frac{1}{r_+} \right) \right)^2
  - \frac{\partial^2}{\partial x^2} +F \left(
    \frac{l(l+1)}{r^2} + m^2 + \frac{F'(r)}{r} \right) \right\} \psi  =0 \]
whose natural conserved quantity
\begin{eqnarray}
\tilde{E} &=& \int_{\Sigma} \left\{ \left| \partial_t \psi \right|^2 + 
\left|
  \partial_{x} \psi \right|^2 + \left( F m^2 + F\frac{l(l+1)}{r^2} + \frac{FF'}{r} - q^2 Q^2 \left( \frac{1}{r}
  - \frac{1}{r_+} \right)^2 \right)
  \vert \psi \vert^2 \right\} \d x \nonumber \\
&=& \int_{\Sigma} \left\{ \left| \left( \partial_t - i\frac{qQ}{r_+}
  \right) \phi \right|^2 + \left|
  \partial_{x} \phi \right|^2  \right. \nonumber \\
&& \left. + \left( F m^2 + F\frac{l(l+1)}{r^2} + \frac{FF'}{r} - q^2 Q^2 \left( \frac{1}{r}
  - \frac{1}{r_+} \right)^2 \right)
  \left| \phi \right|^2 \right\} \d x \nonumber
\end{eqnarray}
is positive definite if and only if $m\geq \left| \frac{qQ}{r_+}
\right|$. Indeed, we have (using $r_- = M - \sqrt{M^2 - Q^2}$ and the fact that $r^2 F= (r-r_-)(r-r_+)$)
\begin{eqnarray*}
Fm^2 - q^2 Q^2 \left( \frac{1}{r} - \frac{1}{r_+} \right)^2 = Fm^2 -
\frac{q^2Q^2 (r-r_+)^2}{r^2r_+^2} &=& Fm^2 -
\frac{q^2Q^2(r-r_+)r^2F}{r^2r_+^2(r-r_-)} \\
&=& F \left[
  m^2 - \frac{q^2Q^2}{r_+^2} \frac{r-r_+}{r-r_-} \right] \, .
\end{eqnarray*}
This quantity is positive in the neighbourhood of $r=+\infty$ if and
only if $m\geq \left| \frac{qQ}{r_+} \right|$. In this case it is positive everywhere outside the black hole,
since $0<\frac{r-r_+}{r-r_-} <1$. All the other terms are clearly
positive except $\frac{FF'}{r}$. However, outside the black hole, $r > r_+ \geq M$, whence
\[ F'(r) = \frac{2M}{r^2} - \frac{2Q^2}{r^3} = 2\frac{Mr -Q^2}{r^3} > 2
  \frac{M^2-Q^2}{r^3} > 0 \, . \qed\]

Equation \eqref{CKGCoord} is of the general form
\begin{equation} \label{ToyModEq}
(\partial_t - i V(x) )^2 \phi - \partial_x^2 \phi + P (x) \phi =0 \, ,
\end{equation}
with
\begin{alignat*}{2}
V(x) & \rightarrow  C_1>0 \mbox{ as } x \rightarrow -\infty \, , & \quad \quad \quad P(x) & \rightarrow  0 \mbox{ as } x \rightarrow -\infty \, , \\
& \rightarrow  0 \mbox{ as } x \rightarrow +\infty \, , & &\rightarrow  C_2 \geq 0 \mbox{ as } x \rightarrow +\infty \, .
\end{alignat*}
A usual toy model for superradiance is obtained by taking equation \eqref{ToyModEq} with potentials $V$ and $P$ that are constant outside of a fixed interval, say $[-L,0]$, $L>0$. A toy model for the massless case is given by
\begin{equation} \label{TMMassless}
P\equiv 0 \, ,~ V \equiv 1 \mbox{ for } x \leq -L \, ,~V \equiv 0 \mbox{ for } x \geq 0 \, ,~ V \mbox{ smooth and decreasing~;}
\end{equation}
whereas for the massive case we choose for $\alpha>0$ and $\beta>0$
\begin{gather}
V \equiv \alpha \mbox{ for } x \leq -L \, ,~V \equiv 0 \mbox{ for } x \geq 0 \, ,~ V \mbox{ smooth and decreasing,} \label{TMMassV} \\
P \equiv 0 \mbox{ for } x \leq -L \, ,~P \equiv \beta \mbox{ for } x \geq 0 \, ,~ P \mbox{ smooth and increasing.} \label{TMMassP}
\end{gather}
The extreme situation often considered as a toy model is the case when $V$ and $P$ as step functions, i.e. $L=0$. We shall see in the numerical study that this changes the situation drastically since hyperradiant modes are then observed.

On the toy model, there is a clear-cut way of measuring the gain of energy, because there is a fixed zone ($\R^+$) where the energy is positive definite. Consequently, taking initial data supported in $\R^+$ and letting them evolve, the gain at time $t$ is given by
\begin{equation} \label{GainToyModel}
G(t) := \frac{E_+ (\phi (t))}{E_+ (\phi (0))} \, ,
\end{equation}
where
\begin{equation}
\label{EnergyComp}
E_+ ( \phi (t) ) := \int_{\R^+} \left( \vert \partial_t \phi (t,x) \vert^2 + \vert \partial_x \phi (t,x) \vert^2 + \beta \vert \phi (t,x) \vert^2 \right) \d x \, .
\end{equation}
We define the asymptotic gain as the following limit when it exists~:
\begin{equation} \label{AsymptoticGainToyModel}
G_\infty  := \lim_{t\rightarrow +\infty} G(t) \, .
\end{equation}
In the physical case of charged Klein-Gordon fields outside a Reissner-Nordstr\o m black hole, for fixed $Q$ and $M$, the region where the potential is negative depends on the physical parameters of the field and can spread out to the whole exterior of the black hole. If we use a similar approach to the one described above for the toy model, we need to change the region of integration for the energy and the localization of the data when we change these parameters. This makes the comparison of gains difficult for different values of $q$, $m$ and $l$ because the same initial data may not be suitable for all cases. Moreover, the extreme situations where the potential is negative everywhere cannot be treated this way. It is much better to find a geometrical way of measuring an energy flux that does not depend on a reference positive definite energy. We shall see when we come to numerical experiments that this geometrical approach is also more convenient for the toy model.

\section{Construction of a geometric conserved quantity} \label{EnergyCurrent}

Our approach is to define a Poynting vector, i.e. an energy current, whose flux across a $t=\, \mbox{constant}$ hypersurface for a given angular momentum $l$ is exactly the conserved energy \eqref{ConservedEnergy}. This is typically done by considering a conserved stress-energy tensor and contracting it with the global timelike Killing vector field. In our case however, the lack of covariance of the equation prevents the existence of such a tensor. Nevertheless, we find a modification of the construction for the Klein-Gordon equation, involving the Maxwell potential, that gives rise to an adequate conserved Poynting vector. We first recall the essential features of Reissner-Nordstr\o m metrics and the definition of the charged wave equation on such backgrounds.

\subsection{Charged scalar fields around Reissner-Nordstr\o m black holes}

An asymptotically flat universe containing nothing but a charged static spherical black hole is described, in Schwarzschild coordinates $(t,r,\theta ,\varphi )$, by the manifold
\[
{\cal M}=\R_{t}\times ]0,\infty[_{r}\times S^{2}_{\theta,\varphi}
\]
equipped with the Reissner-Nordstr{\o}m metric
\begin{gather}
g= g_{\mu \nu} \d x^{\mu} \d x^{\nu} = F(r)\d
t^{2}-F^{-1}(r)\d r^{2}-r^{2}\d \omega^{2}\, , \label{rnmet}\\
\d \omega^{2}=\d \theta^{2} + \sin^{2}\theta \, \d \varphi^{2}\, ,
F(r)=1-\frac{2M}{r}+\frac{Q^{2}}{r^{2}}\, , \nonumber
\end{gather}
where $M> 0$ is the mass of the black hole and $Q$ its charge. Two
types of singularities appear in the expression (\ref{rnmet}) of $g$~:
$\{ r=0\}$ is a true curvature singularity while the spheres where $F$
vanishes, called horizons, are mere coordinate singularities~; they can
be understood as smooth null hypersurfaces by means of Kruskal-type
coordinate transformations (see for example \cite{HaEl}). The fact that these hypersurfaces are null reveals that a
horizon can be crossed one way, but requires a speed greater than the
speed of light to be crossed the other way, hence the name ``event
horizon''. The black hole is the part of space-time situated
beyond the outermost horizon. There are three types of Reissner-Nordstr{\o}m
metrics, depending on the respective importance of $M$ and $Q$.
\begin{enumerate}
\item Sub-extremal case~: for $M>|Q|$, the function $F$ has two real roots
\begin{equation}
r_\pm := M \pm \sqrt{M^2 -Q^2} \, .
\end{equation}
The sphere $\{ r =r_+ \}$ is the outer horizon, or horizon of the black hole and $\{ r=r_- \}$ is the inner or Cauchy horizon.
\item Extreme case~: for $M=|Q|$, $r_+=r_-=M$ is the only root of $F$ and there is only one horizon.
\item Super-extremal case~: for $M< |Q|$, the function $F$ has no real root. There are no horizons, the space-time contains no black hole and the singularity $\{ r=0 \}$ is naked (i.e. not hidden beyond a horizon).
\end{enumerate}
We consider only the sub-extremal case but will mention some striking aspects of the extreme case in this section. On such backgrounds, we study the charged Klein-Gordon equation from the point of view of a distant observer, whose experience of time is described by $t$ and whose perception is limited to the domain of outer communication (i.e. the exterior of the black hole)
\begin{equation} \label{doc}
\doc := \R_t \times \Sigma \, ,~ \Sigma = ]r_+ , +\infty [_r \times S^2_\omega \, .
\end{equation}
The horizon of the black hole is perceived by distant observers as an asymptotic region~: a light ray, or a massive object, falling towards the black hole, will only reach the horizon as $t\rightarrow +\infty$. It is convenient to introduce a new radial coordinate $r_*$, referred to as the Regge-Wheeler tortoise coordinate, such that
\[ \frac{\d r_*}{\d r} = \frac{1}{F} \, .\]
The radial null geodesics travel at constant speed $1$ in this new variable. This will simplify the form of the wave equation as a perturbation of a $1+1$-dimensional wave equation with constant coefficients.

In the case where $M>|Q|$, $r_*$ is given by ($R_0$ being an arbitrary real number)
\begin{eqnarray}
r_* &=& r + M \Log ( r^2 -2Mr + Q^2 ) +
\frac{2M^2 - Q^2}{\sqrt{M^2-Q^2}} \Log \sqrt{\frac{r-r_+}{r-r_-}} +R_0
\nonumber \\
&=& r + \frac{r_+^2}{r_+-r_-} \Log (r-r_+ ) - \frac{r_-^2}{r_+-r_-}
\Log ( r-r_- ) +R_0 \, .\nonumber
\end{eqnarray}
The surface gravity $\kappa_+$ (resp. $\kappa_-$) at the horizon of the black hole (resp. the inner horizon), is defined by
\[ \kappa_+ = F'(r_+) =\frac{r_+-r_-}{r_+^2} ~\left( \mathrm{resp.}~ \kappa_- = F'(r_-) =
  \frac{r_- -r_+}{r_-^2} \right) \, .\]
Thus, the expression of $r_*$ can be further simplified as
\begin{equation} \label{RstarNonExt}
r_* = r +\frac{1}{\kappa_+} \Log ( r-r_+ )+  \frac{1}{\kappa_-} \Log
( r-r_- ) +R_0 \, .
\end{equation}
In the extreme case, we have
\begin{equation} \label{RstarExt}
r_* = r + 2M \Log ( r-M ) - \frac{M^2}{r-M} + R_0 \, .
\end{equation}
In both cases, the map $r\mapsto r_*$ is an analytic diffeomorphism
from $]r_+ , +\infty [$ onto $\R$, $r\rightarrow +\infty$ corresponds
to $r_* \rightarrow +\infty$ and $r\rightarrow r_+$ corresponds to
$r_* \rightarrow -\infty$. Moreover,
\[ \frac{r_*}{r} \longrightarrow 1 \, ,~ \mathrm{as}~r\rightarrow
+\infty \]
and as $r_*\rightarrow -\infty$,
\begin{eqnarray}
\mathrm{for}~M>|Q| \, ,~~~r-r_+ &\simeq & \left( r_+ -r_-
\right)^{\frac{r_-^2}{r_+^2}} e^{-\kappa_+ (r_++R_0)} e^{\kappa_+ r_*} \,
, \nonumber \\
\mathrm{for}~M=|Q| \, ,~~~r-M &\simeq & -\frac{M^2}{r_*} \, .\nonumber
\end{eqnarray}
We see that there is a radical change of behaviour of the function $r(r_*)$ between the sub-extremal case and the extreme case. For the Klein-Gordon equation on these metrics, this means that the mass term $F m^2$ will be exponentially decreasing in $r_*$ at the horizon in the sub-extremal case, whereas in the extreme case it will be a long range term falling-off like $1/r_*$.

In coordinates $(t,r_* ,\theta , \varphi )$, the
Reissner-Nordstr{\o}m metric has the form
\begin{equation} \label{rnmetRstar}
g = F(r) \left( \d t^2 -\d r_*^2 \right) - r^2 \d \omega^2
\end{equation}
and the domain of outer communication is described as
\begin{equation} \label{doc2}
\doc = \R_t \times \Sigma \, ,~\Sigma = \R_{r_*} \times S^2_\omega \, .
\end{equation}
We shall denote by $\Sigma_t$ the level hypersurfaces of the variable $t$ outside the black hole
\begin{equation} \label{Sigma_t}
\Sigma_t = \{ t \} \times \Sigma \, .
\end{equation}
The charged Klein-Gordon equation on a Reissner-Nordstr\o m metric reads
\begin{equation} \label{CKGRN}
\square^A_g f + m^2 f =0 \, .
\end{equation}
where
\[ \square^A_g = (\nabla_a -i q A_a ) ( \nabla^a -i qA^a ) \, , \]
$A_a \d x^a$ being the electromagnetic vector potential (expressed here as
a one-form)
\[ A_a \d x^a = \frac{Q}{r} \d t \, .\]
A short calculation gives the following explicit form of
(\ref{CKGRN})~:
\[ \left\{ F^{-1} \left( \frac{\partial}{\partial t} -i\frac{qQ}{r}
  \right)^2 - \frac{1}{r^2} \frac{\partial}{\partial r} r^2 F
  \frac{\partial}{\partial r} - \frac{1}{r^2} \Delta_{S^2} + m^2 \right\} f =0 \, .\]
Changing the unknown function to $\phi = r f$ simplifies the radial part of the equation and we obtain 
\begin{equation} \label{CKGRNphi}
\left\{ \frac{\partial^2}{\partial t^2} -2i\frac{qQ}{r} \frac{\partial}{\partial t} - \frac{\partial^2}{\partial r_*^2} + F(r) \left(
\frac{-\Delta_{S^2}}{r^2} + m^2 + \frac{F'(r)}{r} \right) - \frac{q^2 Q^2}{r^2} \right\} \phi   =0
\end{equation}
and for a given angular momentum this gives equation \eqref{CKGCoord} with $x = r_*$.

\subsection{Geometric conserved current}

It is well-known that a charged scalar field on a fixed charged background does not admit a conserved stress-energy tensor. This is due to the fact that in this model, the electromagnetic field influences the scalar field but is not influenced in return~; only for the coupled Maxwell-Klein-Gordon system do we have a good conservation of stress-energy. However, if we modify in a simple way the stress-energy tensor of a Klein-Gordon equation to include the electromagnetic vector-potential, we obtain a new non-conserved tensor which has the advantage of inducing a conserved current. The flux of this current through a spacelike slice $\Sigma_t$ for a solution with angular momentum $l$ exactly corresponds to the conserved energy \eqref{ConservedEnergy}.

The stress-energy tensor for the Klein-Gordon equation
\begin{equation} \label{KG}
\square_g u + m^2 u =0 ~ (\mbox{where } \square_g = \nabla_a \nabla^a )\, ,
\end{equation}
is given by
\[ {\cal T}_{ab} = \partial_a u \partial_b u - \frac12 g_{ab} g^{cd} \partial_c u \partial_d u + \frac{m^2}{2} u^2 g_{ab} \]
and satisfies $\nabla^a {\cal T}_{ab} = \left( \square_g u + m^2 u \right) \partial_b u$ which is zero whenever $u$ is a solution of \eqref{KG}. Now if we consider a Killing vector $K^a$, i.e. a vector field whose flow is an isometry, which is characterized by ${\cal L}_K g =0$ or equivalently by the Killing equation
\begin{equation} \label{KillingEq}
\nabla^{(a} K^{b)} =0 \, ,
\end{equation}
the contraction $K_b {\cal T}^{ab}$ gives a conserved current~:
\begin{eqnarray*}
\nabla_a \left( K_b {\cal T}^{ab} \right) &=& K_b \nabla_a {\cal T}^{ab} +  {\cal T}^{ab} \nabla_a K_b \, , \\
&=& {\cal T}^{ab} \nabla_a K_b \\
&=& {\cal T}^{ab} \nabla_{(a} K_{b)} \mbox{ by symmetry of } {\cal T}^{ab} \, ,  \\
&=& 0 \mbox{ by the Killing equation \eqref{KillingEq}.}
\end{eqnarray*}
In particular, taking
\[ K^a \partial_a = \partial_t \, ,\]
we see that the energy current measured by an observer
static at infinity is divergence-free~:
\[ {\cal J}^a := {\cal T}_0^a \, ,~ \nabla_a {\cal J}^a = 0 \, .\]
For our charged, and necessarily complex, Klein-Gordon field, we modify the stress-energy tensor ${\cal T}_{ab}$ as follows. Let us consider the symmetric tensor $T_{ab}$ defined in terms of $f_1 = \Re f$ and $f_2 = \Im f$ by
\begin{equation}
T_{ab} = \sum_{j=1}^2 \left( \partial_a f_j \partial_b f_j  - \frac{1}{2} g^{cd} \left( \partial_c f_j \partial_d f_j +  q^2 A_c A_d \left( f _j\right)^2 \right) g_{ab} + \frac{m^2}{2} (f_j )^2 g_{ab} \right)  \, .\label{Stress-Energy}
\end{equation}
We use it to construct a Poynting vector by contracting $T_{ab}$ with $\partial_t$
\begin{eqnarray}
J^a \partial_a := T_0^a \partial_a &=& \sum_{j=1}^2 \left( \partial_t f_j \nabla f_j + \frac12 \left[ -\langle \nabla f_j , \nabla f_j \rangle_g - q^2 \langle A , A \rangle_g f_j^2 + m^2 f_j^2 \right] \partial_t \right) \nonumber \\
&=& \sum_{j=1}^2 \left( \partial_t f_j \nabla f_j + \frac12 \left[ -\langle \nabla f_j , \nabla f_j \rangle_g + (m^2 - F^{-1} \frac{q^2Q^2}{r^2} ) f_j^2 \right] \partial_t \right) \, .\label{Poynting}
\end{eqnarray} 
The following lemma summarizes the properties of $T_{ab}$~; its proof is given in appendix \ref{ProofLemma}.
\begin{lemma} \label{LemmaConsLawT}
The tensor $T_{ab}$ does not satisfy $\nabla^a T_{ab}=0$, but  the Poynting vector field $J^a = T^a_0$ is divergence-free~:
\begin{equation} \label{ConsLawSET}
\nabla^a J_{a} = 0 \, .
\end{equation}
\end{lemma}
\begin{remark}
We shall establish in the next section that the flux of $J^a$ across a hypersurface $\Sigma_t$ is exactly the conserved energy \eqref{ConservedEnergy} for a given angular momentum $l \in \N$.
\end{remark}

\subsection{Energy fluxes across two types of hypersurfaces}

Given $S$ a piecewise ${\cal C}^1$ oriented hypersurface, $l^a$ a vector field transverse to $S$ compatible with the orientation of $S$, $n^a$ a normal vector field to $S$ such that $l_a n^a =1$, the flux of a vector field $V^a$ across $S$ reads
\[ {\cal F}_{S} (V) = \int_{S} V_a n^a ( l \lrcorner \dvol ) \]
and is independent of the choice of $l^a$ and $n^a$ satisfying the above properties. For the flux of the energy current $J^a\partial_a$ across a hypersurface $\Sigma_t$, we may choose $n = F l = \partial_t$ and the flux is therefore given by
\begin{eqnarray}
{\cal F}_{\Sigma_t} (J) &=& \frac12 \sum_{j=1}^2 \int_{\Sigma_t} \left( (\partial_t f_j )^2 + (\partial_{r_*} f_j )^2 + \frac{F}{r^2} \vert \nabla_{S^2} f_j \vert^2 \right. \nonumber \\
&& \hspace{0.5in} \left. + \left( Fm^2 - \frac{q^2Q^2}{r^2} \right) (f_j )^2 \right) r^2 \sin \theta \d r_* \wedge \d \theta \wedge \d \varphi \nonumber \, ,\label{EnSigmatf}
\end{eqnarray}
where
\[ \vert \nabla_{S^2} f_j \vert^2 = (\partial_\theta f_j )^2 + \frac1{\sin^2 \theta} (\partial_\varphi f_j)^2 \, .\]
Across a hypersurface
\[ [t_1 , t_2 ] \times S_R \, ,~ S_R = \{ R \}_r \times S^2_{\omega} \, ,\]
orientated  by $\partial_r$, we can take $l = -F^{-1}n = \partial_r$ and we get
\[ {\cal F}_{{[t_1 , t_2 ] \times S_R}} (J) = \int_{[t_1 , t_2 ] \times S_R} \left( -\partial_t f_1 \partial_{r_*} f_1 - \partial_t f_2 \partial_{r_*} f_2 \right) ( -r^2 \sin \theta \d t \wedge \d \theta \wedge \d \varphi ) \]
since $\partial_r \lrcorner \dvol = - r^2 \sin \theta \d t \wedge \d \theta \wedge \d \varphi$ and $F \partial_r = \partial_{r_*}$.
We give the expressions of these fluxes in terms of $\phi_1 = rf_1$ and $\phi_2 = rf_2$. The energy flux across $\Sigma_t$ becomes
\begin{eqnarray*}
{\cal F}_{\Sigma_t} (J) &=& \frac12 \sum_{j=1}^2 \int_{\Sigma_t} \bigg\{ (\partial_t \phi_j )^2 + \left( \partial_{r_*} \phi_j - \frac{F}{r} \phi_j \right)^2 + \frac{F}{r^2} \vert \nabla_{S^2} \phi_j \vert^2 \nonumber \\
&& \hspace{0.5in} + \left( Fm^2 -\frac{q^2Q^2}{r^2} \right) (\phi_j )^2 \bigg\} \sin \theta \d r_* \wedge \d \theta
\wedge \d \varphi \, .
\end{eqnarray*}
Developing the second term for $j=1$, we get
\begin{gather*}
\int_{\Sigma_t}  \left( \partial_{r_*} \phi_1 - \frac{F}{r} \phi_1 \right)^2 \sin \theta \d r_* \wedge \d \theta
\wedge \d \varphi \\
= \int_{\Sigma_t}  \left( \left(\partial_{r_*} \phi_1 \right)^2 + \frac{F^2}{r^2} (\phi_1)^2 - 2 \frac{F}{r} \phi_1 \partial_{r_*} \phi_1 \right) \sin \theta \d r_* \wedge \d \theta \wedge \d \varphi \\
= \int_{\Sigma_t}  \left( \left(\partial_{r_*} \phi_1 \right)^2 + \frac{FF'}{r} (\phi_1)^2 - \partial_{r_*} \left( \frac{F}{r} \phi_1^2 \right) \right) \sin \theta \d r_* \wedge \d \theta \wedge \d \varphi \, ;
\end{gather*}
the last term vanishes for smooth compactly supported functions and therefore also for functions in the finite energy Hilbert space by a standard density argument. This entails the following expression of the energy flux through $\Sigma_t$~:
\begin{eqnarray*}
{\cal F}_{\Sigma_t} (J) &=& \frac12 \int_{\Sigma_t} \left( \vert \partial_t \phi \vert^2 + \vert \partial_{r_*} \phi \vert^2  + \frac{F}{r^2} \vert \nabla_{S^2} \phi \vert^2 \right. \\
&& \hspace{1in} \left. + \left( \frac{FF'}{r} +Fm^2- \frac{q^2Q^2}{r^2} \right) \vert \phi \vert^2 \right) \sin \theta \d r_* \d \theta  \d \varphi
\end{eqnarray*}
which, once restricted to an angular momentum $l$, is exactly the expression (\ref{ConservedEnergy}) of the conserved energy $E(t)$.

The outgoing energy flux across $[t_1 , t_2 ] \times S_R$ in terms of $\phi$ has the form
\begin{eqnarray}
{\cal F}_{[t_1 , t_2 ] \times S_R} &=& \sum_{j=1}^2 \int_{[t_1 , t_2 ] \times S_R} \left( - \partial_t \phi_j \left( \partial_{r_*} \phi_j - \frac{F}{r} \phi_j \right) \right) (- \sin \theta \d t \wedge \d \theta \wedge \d \varphi ) \nonumber \\
&=& - \sum_{j=1}^2 \int_{[t_1 , t_2 ] \times [0,\pi ] \times [0, 2\pi [ } \left( \partial_t \phi_j \left( \partial_{r_*} \phi_j - \frac{F}{r} \phi_j \right) \right) \sin \theta \d t \d \theta \d \varphi \, , \label{FluxRSortant}
\end{eqnarray}
since $l \lrcorner \dvol = - \sin \theta \, \d t \wedge \d \theta \wedge \d \varphi$ gives the right orientation.

\subsection{Practical measure of energy gain}

Given initial data belonging to a subspace on which the energy is positive definite, $R>0$ large enough so that the support of the initial data is entirely contained in $\{r_* < R \}$, the quantity we wish to measure is the ratio ${\cal G}_R(t)$ of the total energy radiated away through $S^2_R$ at time $t$, to the energy of the data. We shall observe superradiance if we can find such initial data for which, as $t$ becomes large, this ratio stabilizes to a value that is strictly larger than $1$. With our previous construction of conserved energy current, ${\cal G}_R$ is easy to evaluate for any solution whose initial data are compactly supported~; it is given as follows
\begin{equation} \label{EnergyGain}
{\cal G}_R (t) = \frac{1}{{\cal F}_{\Sigma_0}} {\cal F}_{{[0,t] \times S^2_R}} \, .
\end{equation}
The asymptotic gain, which is the theoretical quantity of interest here, is defined as the following limit, if it exists~:
\begin{equation} \label{AsymptoticGain}
{\cal G}_\infty := \lim_{t\rightarrow +\infty} {\cal G}_R (t) \, .
\end{equation}

Note that it is not at all obvious that this limit should exist. The fact that ${\cal G}_\infty$ is well-defined and independent of $R>0$ chosen as above is a scattering property that we observe numerically.

\section{Numerical study of superradiance} \label{Numerics}

We now use numerical tools in order to observe superradiance for the toy-model as well as for the Reissner-Nordstr\o m model.

\subsection{Discretization of the problem}

Equation \eqref{ToyModEq} can be expressed as the following 
first-order system
\begin{equation}
\partial _t \left(
\begin{array}{l}
u\\v
\end{array}
 \right)
-
\left( 
\begin{array}{cc}
iV & 1\\ \partial _x^2 -P & iV
\end{array}
\right)
 \left(
\begin{array}{l}
u\\v
\end{array}
 \right)
=
 \left(
\begin{array}{l}
0\\0
\end{array}
 \right)
\label{syst}
\end{equation}
with $u:=\phi$ and $v:=(\partial _t - i V)u$. 
This is a linear evolution system that can
be numerically solved using a finite difference scheme both in time and 
space. We first consider a semi-implicit time discretization 
of the problem~: $\delta t$ denoting the prescribed timestep, 
we set $t_n=n\,\delta t$, $u^n = u(t_n,.)$, $v^n=v(t_n,.)$ and 
write that (\ref{syst}) is satisfied
at the midpoint time $t_{n+1/2}=(t_n+t_{n+1})/2$. We respectively
approximate time derivatives $\partial_t u$ and $\partial_t v$
with $(u^{n+1}-u^n)/\delta t$ and $(v^{n+1}-v^n)/\delta t$. Expressing
$u(t_{n+1/2},.)$ and $v(t_{n+1/2},.)$ with the second order approximations
$(u^n+u^{n+1})/2$ and $(v^n+v^{n+1})/2$, (\ref{syst}) reduces to 
the simplified system
\begin{gather*}
\left( 
\begin{array}{cc}
\displaystyle
(1 - iV \frac{\delta t}{2}) \mathrm{Id}  & 
\displaystyle-\frac{\delta t}{2}\mathrm{Id} \\
\vspace*{-0.2cm}\\
 -\displaystyle
\frac{\delta t}{2}(\partial _x^2-P) & 
\displaystyle
(1 - iV \frac{\delta t}{2}) \mathrm{Id} 
\end{array}
\right)
 \left(
\begin{array}{l}
u^{n+1}\\ \\ \\v^{n+1}
\end{array}
 \right) \hspace{2in} \\
\hspace{1in}=
\left( 
\begin{array}{cc}
\displaystyle
(1 + iV \frac{\delta t}{2}) \mathrm{Id} & 
\displaystyle \frac{\delta t}{2}\mathrm{Id}\\
\vspace*{-0.2cm}\\
\displaystyle
\frac{\delta t}{2}(\partial _x^2-P) & 
\displaystyle
(1 + iV \frac{\delta t}{2}) \mathrm{Id}
\end{array}
\right)
 \left(
\begin{array}{l}
u^{n}\\ ~\\~\\v^{n}
\end{array}
 \right) ,
\end{gather*}
``$\mathrm{Id}$'' denoting the Identity matrix. The space operator $\partial_x^2$ is then discretized using the standard
$3$ point approximation as follows~: when considering a uniform spatial mesh 
with space step $h$, i.e. $x_j = x_0+j\, h$ for all $j$, we have that for any function $f$ depending on $x$
\begin{equation}
\partial _x^2 f(x_j)\approx\frac{1}{h^2}(f(x_{j+1})-2f(x_j)+f(x_{j-1})) \, ,
\label{DiscDiff}
\end{equation}
up to a second order error term. Dealing with a bounded
numerical domain, say $[-L,L]$, we are faced with unknowns 
$U^n=(u_1^n,\ldots,u_J^n)^T\in \C^J$ and $V^n=(v_1^n,\ldots,v_J^n)^T\in 
\C^J$.
This leads to the following finite dimensional linear
system to solve at each iteration~:
$$\left( 
\begin{array}{rr}
A & -B\\
 -C & A
\end{array}
\right)
 \left(
\begin{array}{l}
U^{n+1} \\V^{n+1}
\end{array}
 \right)
=
\left( 
\begin{array}{rr}
D & B\\
C & 
D
\end{array}
\right)
\left(
\begin{array}{l}
U^{n}\\V^{n}
\end{array}
 \right).
$$ 
with $A = (1 - iV \frac{\delta t}{2}) \mathrm{Id}$, $B=\frac{\delta t}{2}\mathrm{Id}$,
$C=\frac{\delta t}{2}(\Delta_2-P)$ and $D = (1 + iV \frac{\delta t}{2}) \mathrm{Id}$, where $\Delta_2$ stands for the tridiagonal
matrix arising when considering the discretization of $\partial_x^2$
given by (\ref{DiscDiff}).

This scheme is known to be convergent under the assumption that the 
Courant-Friedrichs-Lewy condition $|\delta t/h|\le 1$ is fulfilled~: this grants the numerical stability of the scheme and the
Lax-Richtmyer convergence Theorem then gives the convergence (see for example
\cite{strik} for technical details of the numerical analysis of 
finite differences schemes). This means
that taking well-adapted time and space meshgrid parameters leads to 
an accurate approximation of the exact
solution of the wave equation. For all the tests presented in this paper, 
we have checked that the numerical solutions that are obtained do not
depend on the choice of the grid parameters.

\subsection{The boundary conditions}

Since we restrict ourselves to a bounded spatial domain, often referred to 
as the {\it computational domain}, we have to be very
careful with the  numerical treatment of the boundary. Indeed, for
sufficiently large times, the solution reaches the frontier
and ill-adapted boundary conditions may lead to unphysical solutions
that cannot represent the restriction of the correct solution to the 
domain under consideration. As a consequence, waves may be reflected by
the boundary and propagate in the wrong direction.
This problem occurs even for the homogeneous wave equation
\begin{equation}
\partial_t^2 \phi- \partial_x^2 \phi=0
\label{Wave}
\end{equation}
that admits the general solution 
$\phi(t,x)=F(x-t)+G(x+t)$ involving two classes of solutions that propagate at speeds $\pm 1$.
In this case, it is possible to derive \emph{transparent} conditions that 
enable to compute a numerical solution that will not be affected by
the boundary. 
In the last three decades, there has been a sustained effort to extend these boundary conditions
to more general equations. Transparent conditions are delicate to use in the
general case, since they involve nonlocal operators as well as possibly infinite expansions. 
For a comprehensive study of these conditions
as well as the search for approximate conditions, one can refer to the
pioneering works \cite{enma}, \cite{haltr} and \cite{lind} (note that
such conditions have been studied for other problems such as diffusion or Schr\"odinger equations, for which propagation at finite velocity no 
longer holds). It can be shown that for the model equation
\begin{equation}
(\partial_t - iV(x))^2 \phi-\partial _x^2 \phi=0\, ,
\label{mod}
\end{equation}
the transparent condition that has to be considered at the right extremity of
the segment simply reads 
$$
(\partial_t - iV(x)) \phi+ \partial _x \phi=0\, .
$$
This is a local condition that can easily be implemented in the finite difference
discretization. For $V \equiv 0$, one recognizes the condition $\partial_t \phi + \partial_x \phi =0$ which selects propagation at speed $+1$ across the right boundary. The symmetric condition
\[ (\partial_t - iV(x)) \phi- \partial _x \phi=0 \]
is set at the left boundary. The expression of these boundary conditions can be obtained by means of the algebraic identity $A^2 - B^2 = (A-B)(A+B)$ which holds when $A$ and $B$ are differential operators.

We now illustrate the influence of the boundary condition on the
calculated solution, for several values of constant potentials $V$ and $P$. In all our simulations, the timestep 
$\delta t$ and the spatial mesh $h$ are taken in order to fulfill
the Courant-Friedrichs-Lewy condition $\delta t/h\le 1$ that grants the numerical stability and the convergence of the scheme. 
We first perform computations for the homogeneous case $P=V=0$~: we deal with the wave equation (\ref{Wave}) for initial data 
$$
\phi(0,x)=\varphi(x)\, , \quad \partial_t \phi(0,x)=\varphi'(x) \, , \quad
x\in \R \, ,
$$
where $\varphi(x)=e^{-x^2}$. The associated solution solves the transport equation
$\partial_t \phi - \partial_x \phi=0$ and propagates at constant speed $-1$~; it is an incoming wave packet with zero frequency.
We performed our
computations on the space domain $[-5,5]$ discretized with $h=0.04$, 
until final computational time $T=10$. 

\begin{figure}[htb]
\centering
  \begin{tabular}{@{}ccc@{}}
   \hspace{-0.1in}
   \includegraphics[width=.33\textwidth,height=6.cm]{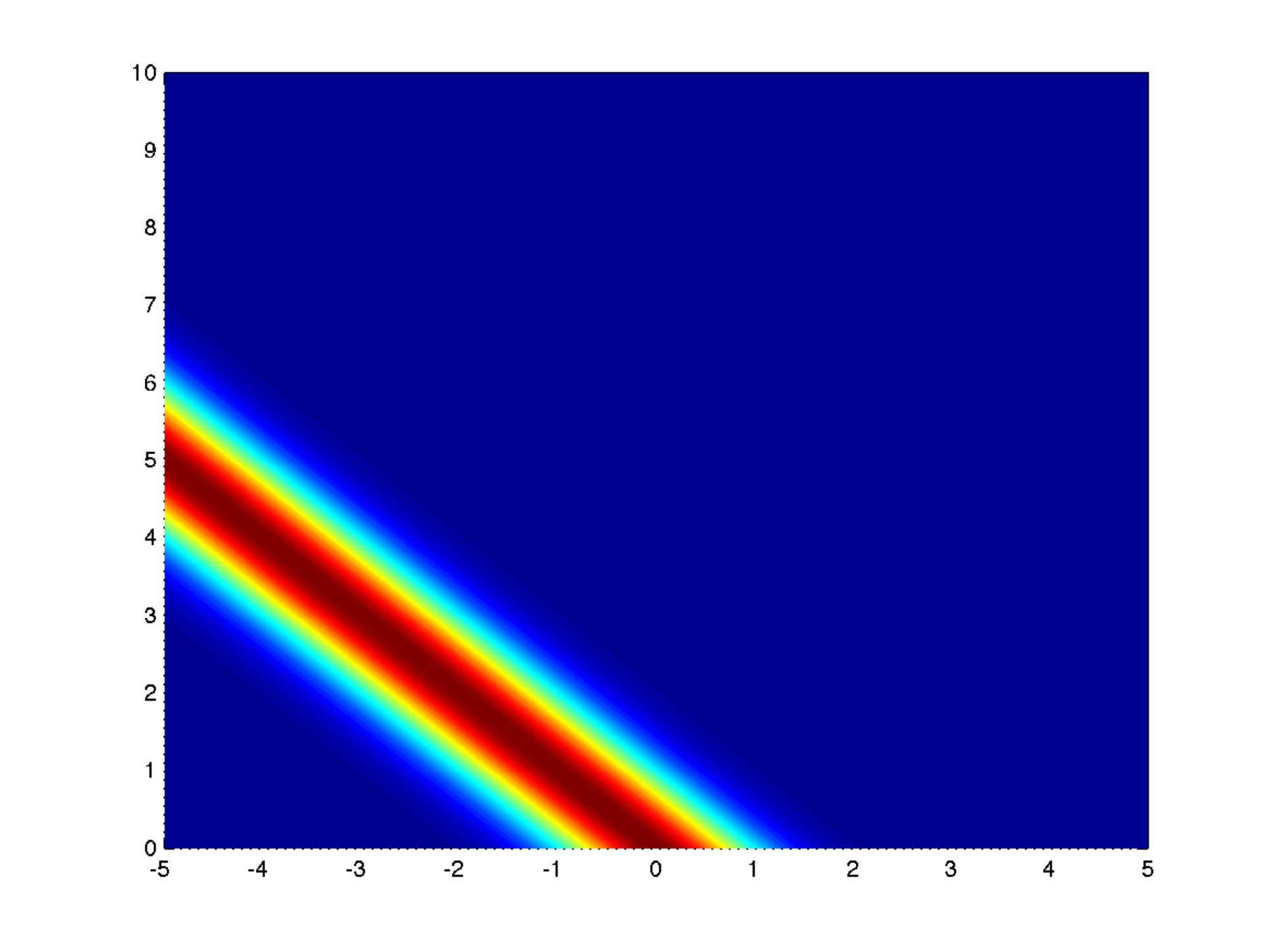} &
   \hspace{-0.2in}
   \includegraphics[width=.33\textwidth,height=6.cm]{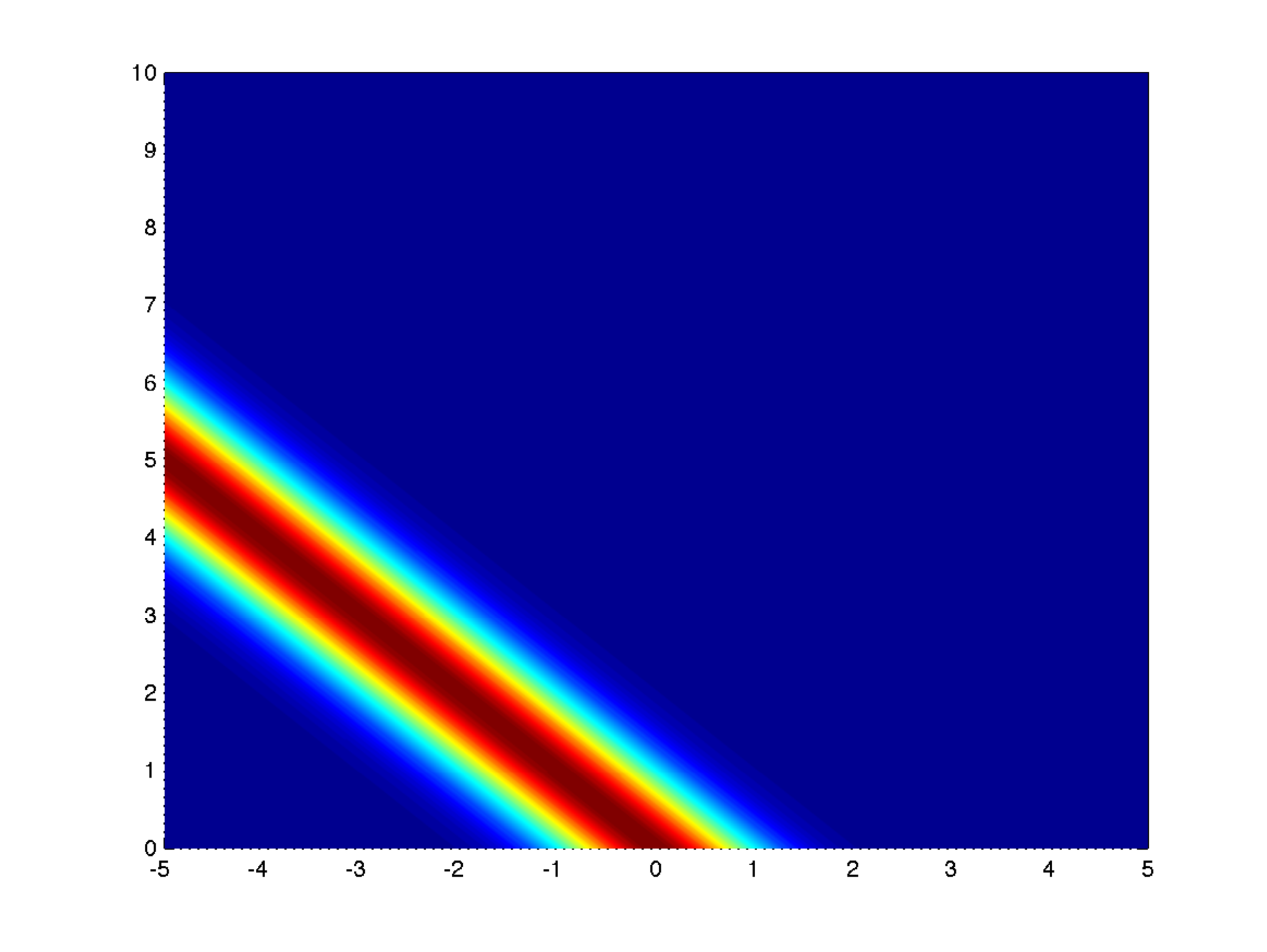} &
   \hspace{-0.2in}
   \includegraphics[width=.33\textwidth,height=6.cm]{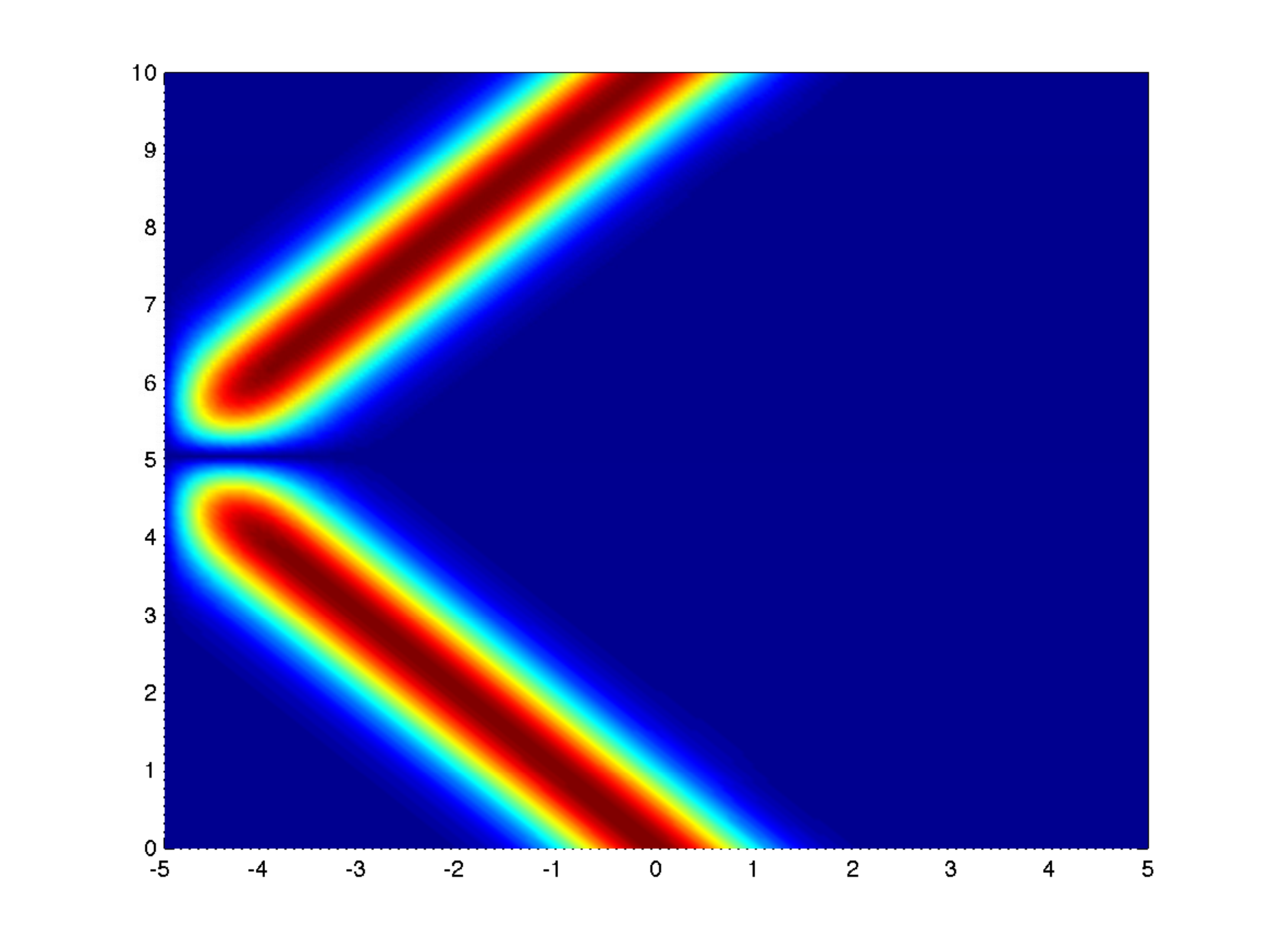}
  \end{tabular}
\caption{Solution computed first with transparent boundary conditions, second on a larger spatial domain and shown only on the smaller domain, third with Dirichlet boundary conditions.}\label{Fig123}
\end{figure}

As shown in Figure \ref{Fig123}, the
transparent condition enables the solution to go outside the numerical
domain and the boundary has no effect on long-time dynamics. The computed
solution mimics the profile of the one calculated on a larger domain, whereas
taking Dirichlet homogeneous conditions $\phi=0$ on the boundary 
gives birth to a reflected wave that propagates to the right. 

\begin{figure}[htb]
\centering
  \begin{tabular}{@{}ccc@{}}
   \hspace{-0.1in}
   \includegraphics[width=.33\textwidth,height=6.cm]{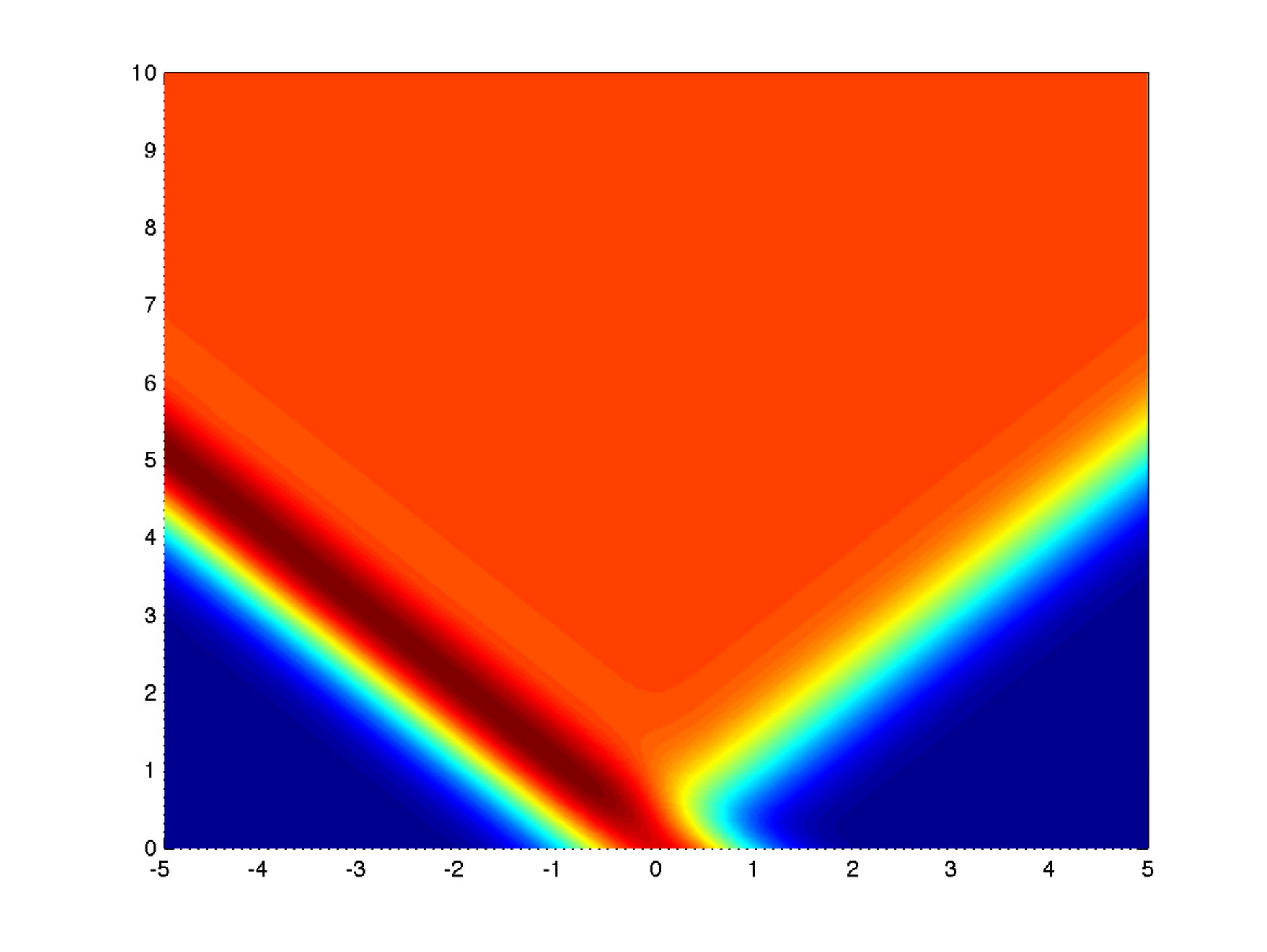} &
   \hspace{-0.2in}
   \includegraphics[width=.33\textwidth,height=6.cm]{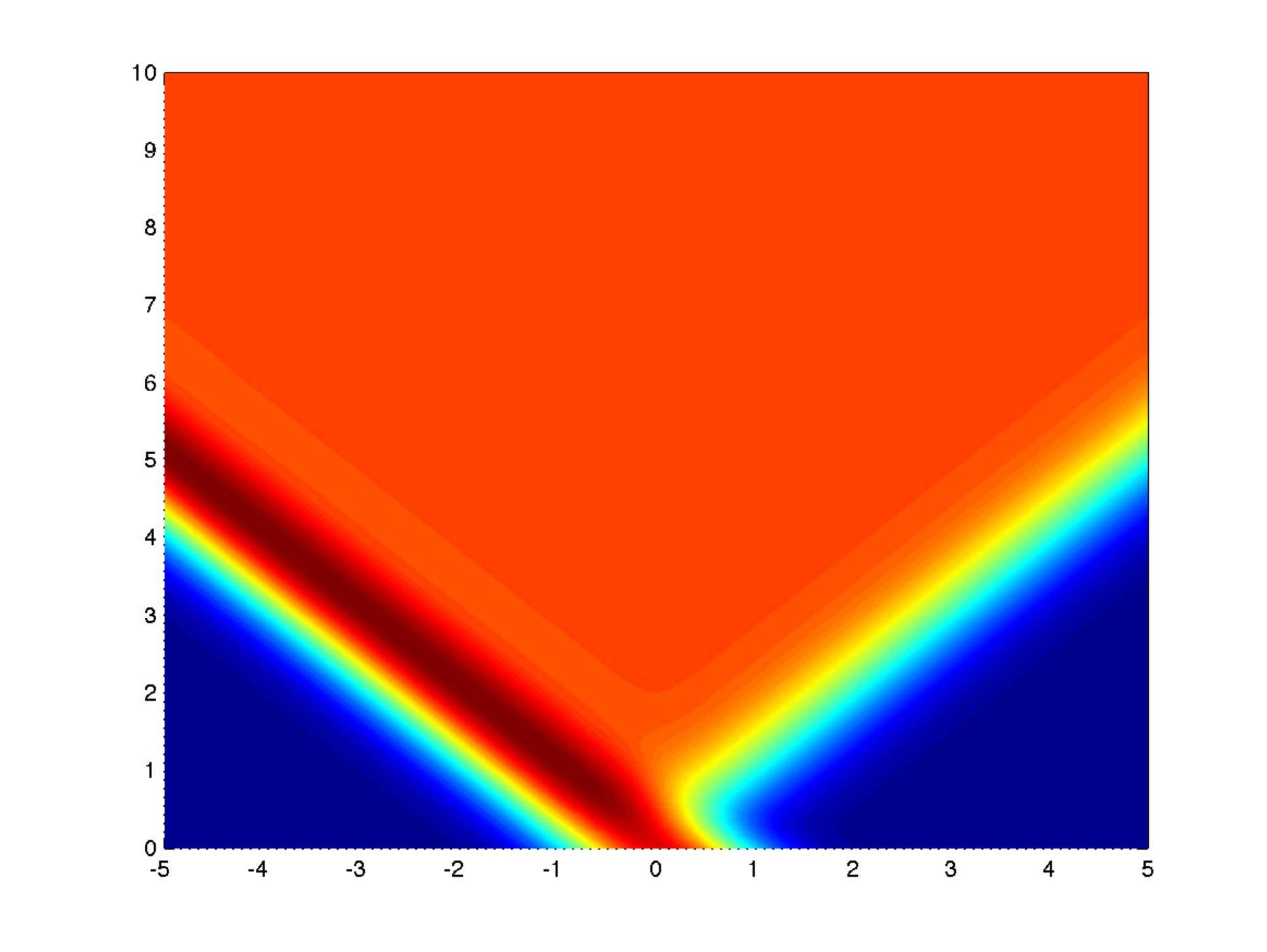} &
   \hspace{-0.2in}
   \includegraphics[width=.33\textwidth,height=6.cm]{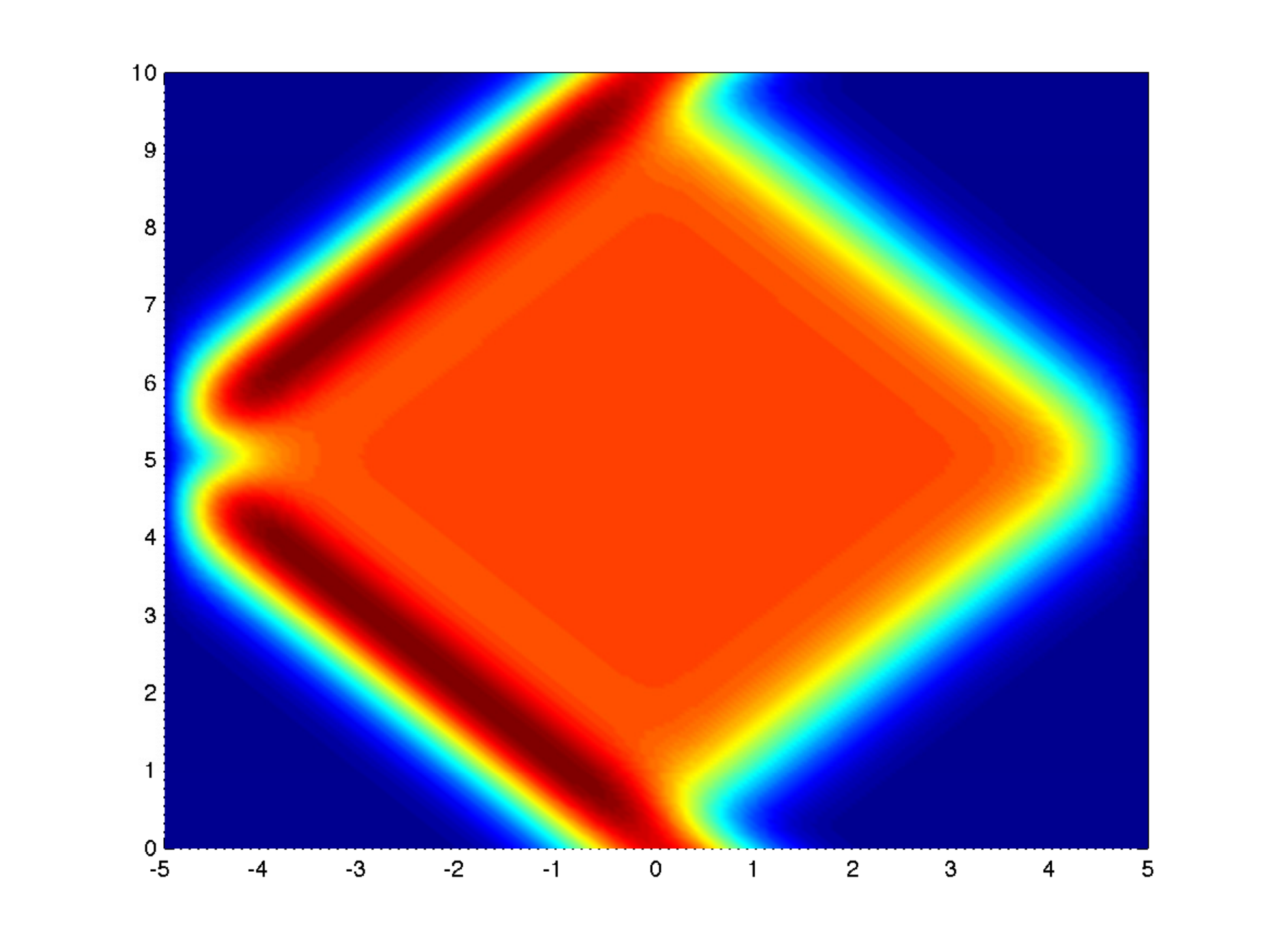}
  \end{tabular}
\caption{Amplitude of the solution computed first with transparent boundary conditions, second on a larger domain and shown only on the smaller domain, third with Dirichlet boundary conditions.} \label{Fig456}
\end{figure}

\begin{figure}[htb]
\centering
  \begin{tabular}{@{}ccc@{}}
   \hspace{-0.1in}
   \includegraphics[width=.33\textwidth,height=6.cm]{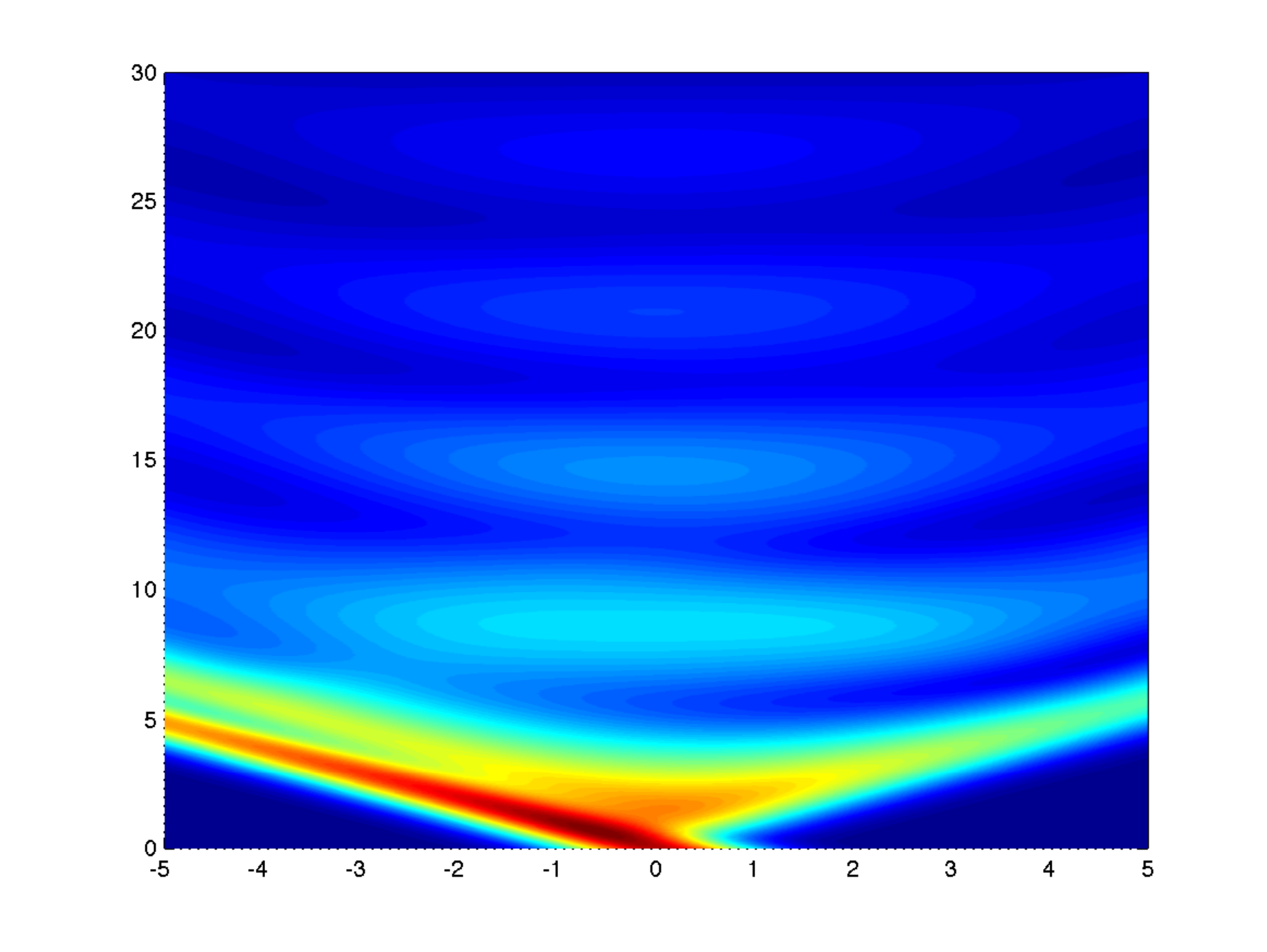} &
   \hspace{-0.2in}
   \includegraphics[width=.33\textwidth,height=6.cm]{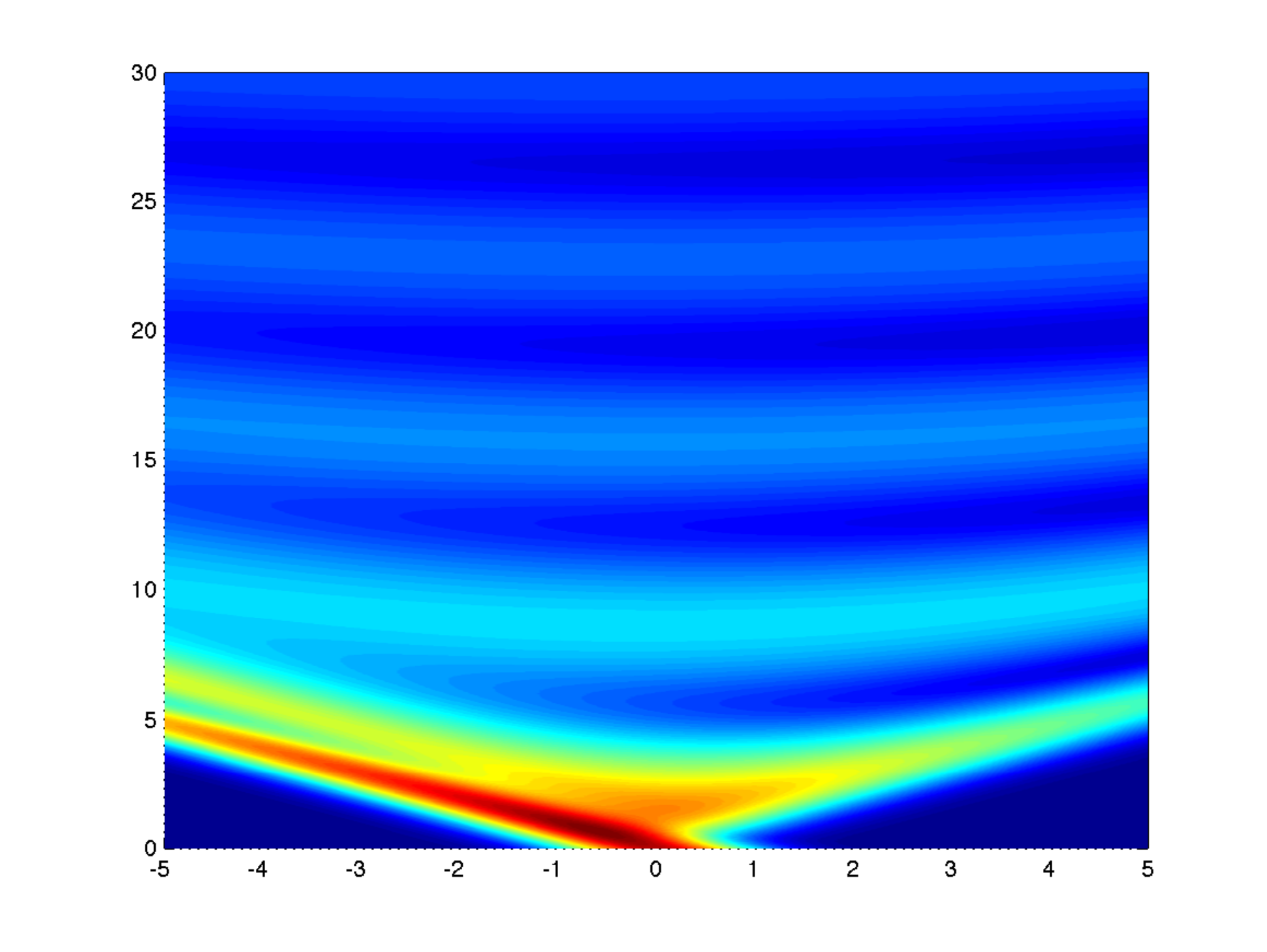} &
   \hspace{-0.2in}
   \includegraphics[width=.33\textwidth,height=6.cm]{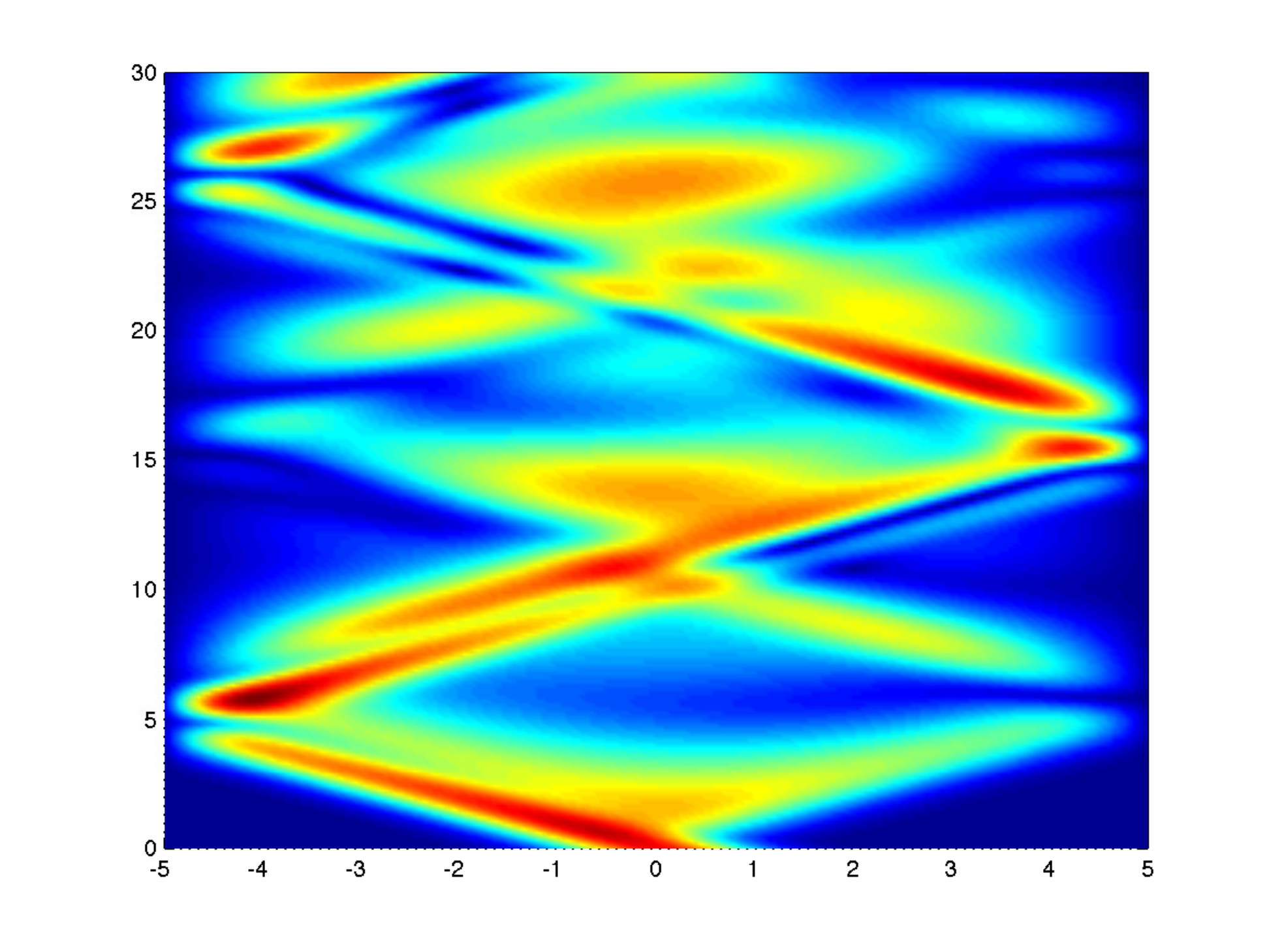}
  \end{tabular}
\caption{Amplitude of the solution computed first with transparent boundary conditions, second on a larger domain and shown only on the smaller domain, third with Dirichlet boundary conditions.} \label{Fig789}
\end{figure}

We now perform the same computations for constant potentials $V\equiv 1$ and
$P\equiv 0$, using the same computational parameters as before and starting from the same
Cauchy data. The same conclusions can be drawn, as can be seen in 
Figure \ref{Fig456} where the amplitude is now plotted, 
since the solution is now complex-valued~: once again, the computed
solution is not affected by the boundaries. Let us mention here that as opposed to the
homogeneous case, the support of the solution spreads out through time and the transparent
conditions at both extremities are required, even for incoming wave packet Cauchy data.  
In the case $P\ne 0$, the problem becomes more difficult to handle~: the
transparent boundary condition is much more costly to
implement, since it involves the evaluation of a pseudodifferential operator that is non local in time. Indeed the factorization
\[ A^2 - B^2 +C = (A- (B^2-C)^{1/2}) (A+ (B^2-C)^{1/2}) \]
shows that the boundary condition now involves an operator with a square root symbol. 
Nonlocal boundary conditions are delicate to implement because first,
they require the full storage of the solution at the boundary from the initial computational time until current time and also, the computation of the integral
boundary operator may lead to an additional error. For this reason, we prefer
to deal with a splitting strategy in order to get rid of this difficulty and 
to use local exact transparent conditions for the resolution of the homogeneous
equation (that is for $P=0$).
Between two consecutive time increments, the
initial problem \eqref{ToyModEq} is decomposed into the two elementary problems
$$
(\partial_t - iV(x))^2 \phi- \partial^2 _x \phi=0
\quad \mbox{and} \quad
\partial_t^2 \phi + P \phi=0.
$$
This first equation is numerically solved as above
with the use of the transparent conditions, whereas the second one that
reduces to an ODE is integrated using a classical implicit scheme.
Simulations have been performed with the same discretization 
parameters as before, for the case $P=0.2$ and $V=1$ and up to final time $T=30$. The classical
second-order Strang splitting (see \cite{strang})
is used in order to get a global
second-order method in time. The results displayed in Figure \ref{Fig789} show that once again transparent conditions
produce a correct approximation of the solution on a larger domain (on which the boundary has no influence), even if small 
differences with the reference solution can be observed as a consequence of the use of 
the splitting algorithm.
The problem of the boundary treatment is crucial for the
computation of the energy gain at large times when seeking
superradiance evidence, as will be seen later.

\subsection{Numerical results for the toy model}

We now study the toy model \eqref{ToyModEq} for potentials $V$ and $P$ 
defined as follows~:
$$
V(x) = \left\{
\begin{array}{ll}
\alpha & \mbox{if~} x\le -L \, ,\\
\frac{\alpha}{2} \Big(1-\sin(\frac{\pi}{L} (x+L/2))\Big)
&\mbox{if } -L \leq x \leq 0 \, ,\\ 
0 & \mbox{if~} x\ge 0 \, .
\end{array}
\right.
$$

Such a potential $V$ is a smooth approximation of the limit case 
$V(x)=\alpha H(-x)$, where $H$ stands for the Heaviside function. Once $V$
is prescribed, we set $P=\beta(1-V/\alpha)$. Using such a relation,
we have that $V(x)\equiv \alpha$, $P(x)\equiv 0$ for $x\le -L$ and
$V(x)\equiv 0$, $P(x)\equiv \beta$ for $x\ge 0$. 

Once again, we consider incoming wave packet initial data
\begin{eqnarray}
\phi (0,x) &=& e^{i \omega x} e^{-(x-x_0)^2} \, , \label{DataToyU0} \\
\partial_t \phi (0,x) &=& \partial_x \phi (0,x) = (i \omega -2 (x-x_0)) e^{i \omega x} e^{-(x-x_0)^2} \, , \label{DataToyU1}
\end{eqnarray}
where $x_0\gg 1$ is fixed and the frequency $\omega$ can be varied to 
observe the high and low energy behaviours. We analyze the behaviour of the solution of equation \eqref{ToyModEq} associated with such data. The gain is measured using formula \eqref{GainToyModel}.

\begin{figure}[htb]
\centering
  \begin{tabular}{@{}cc@{}}
   \hspace{-0.2in}
   \includegraphics[width=.5\textwidth]{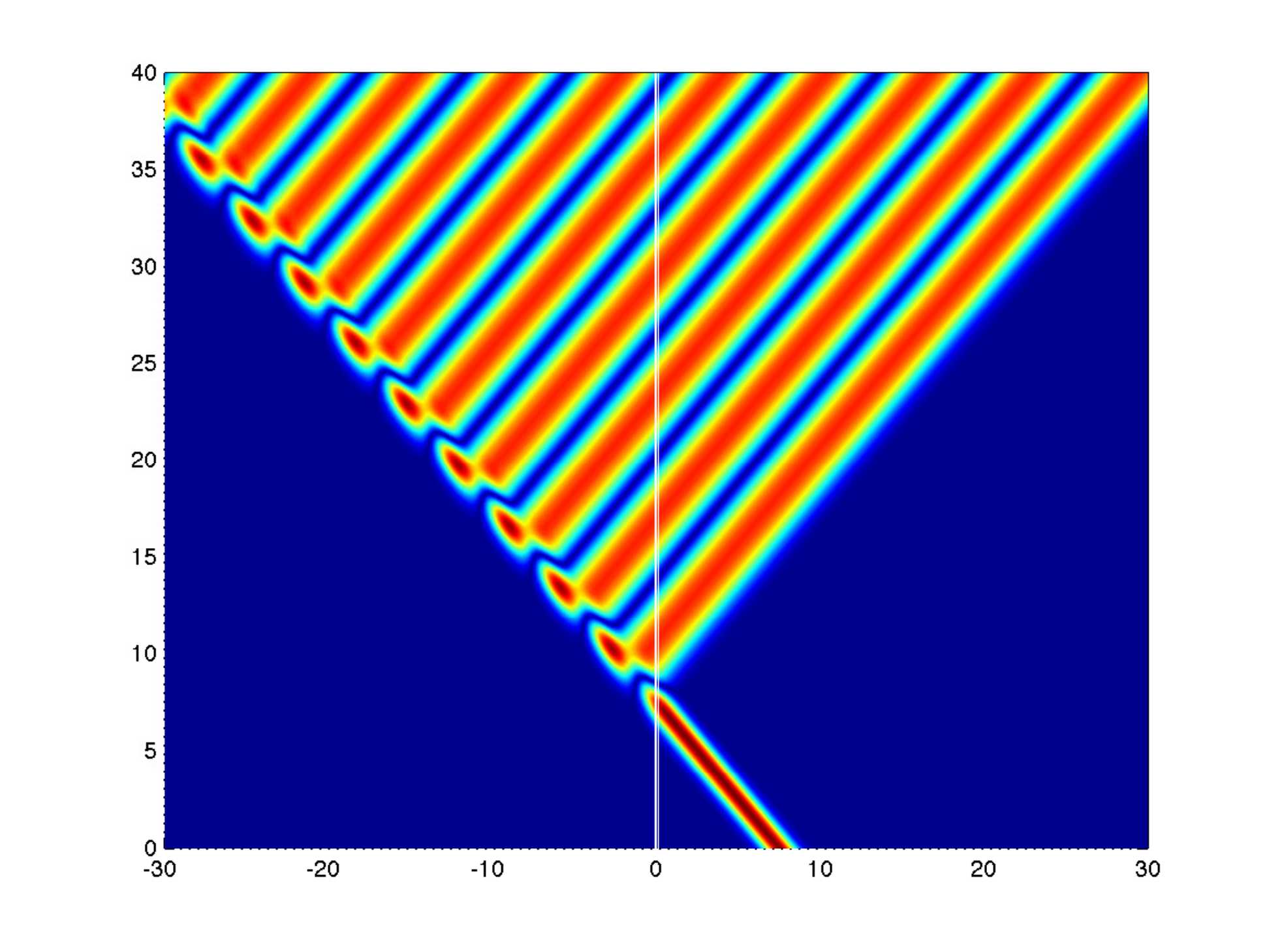} &
   \hspace{-0.4in}
   \includegraphics[width=.5\textwidth]{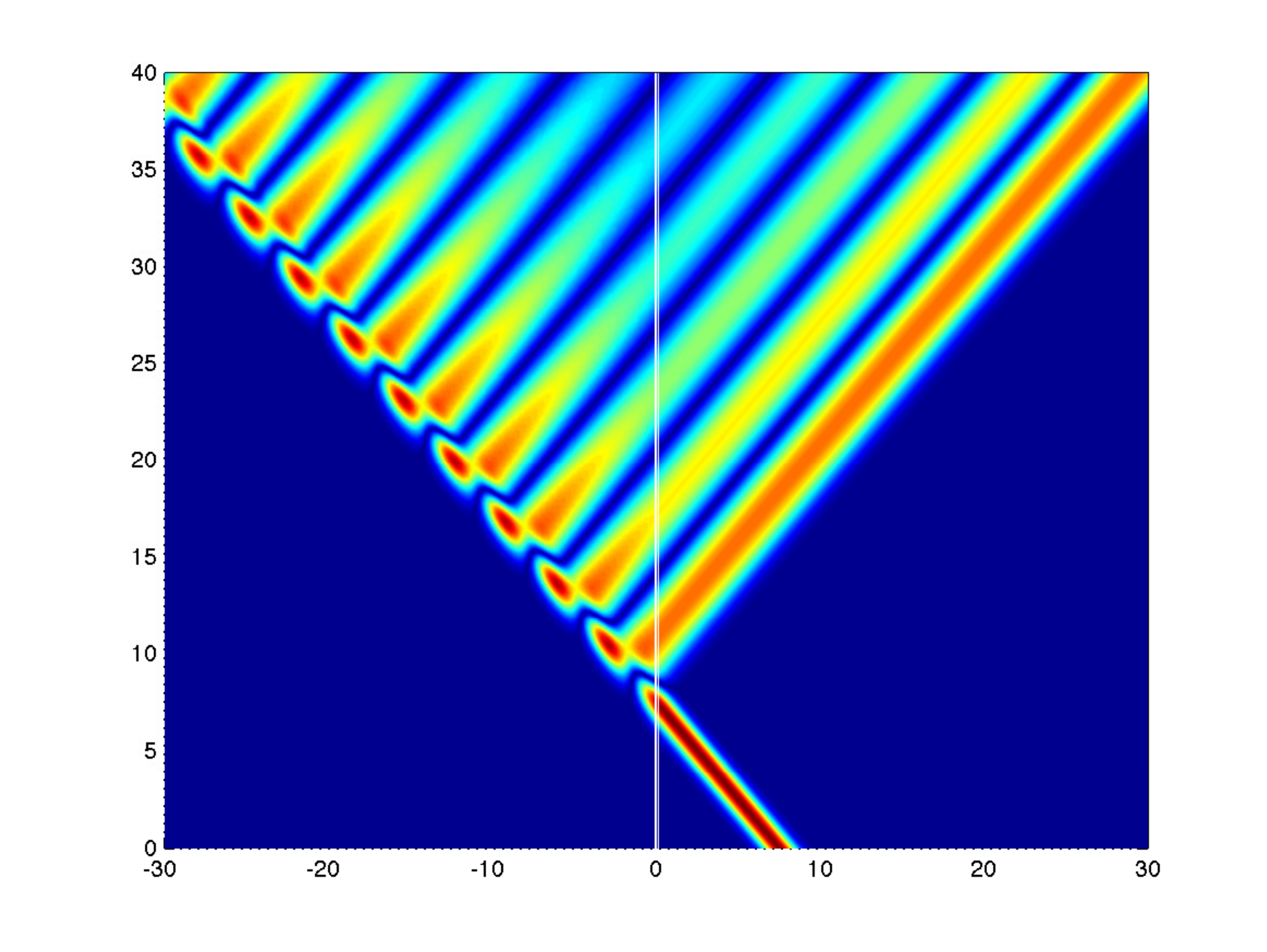} \\
   \hspace{-0.2in}
   \includegraphics[width=.5\textwidth]{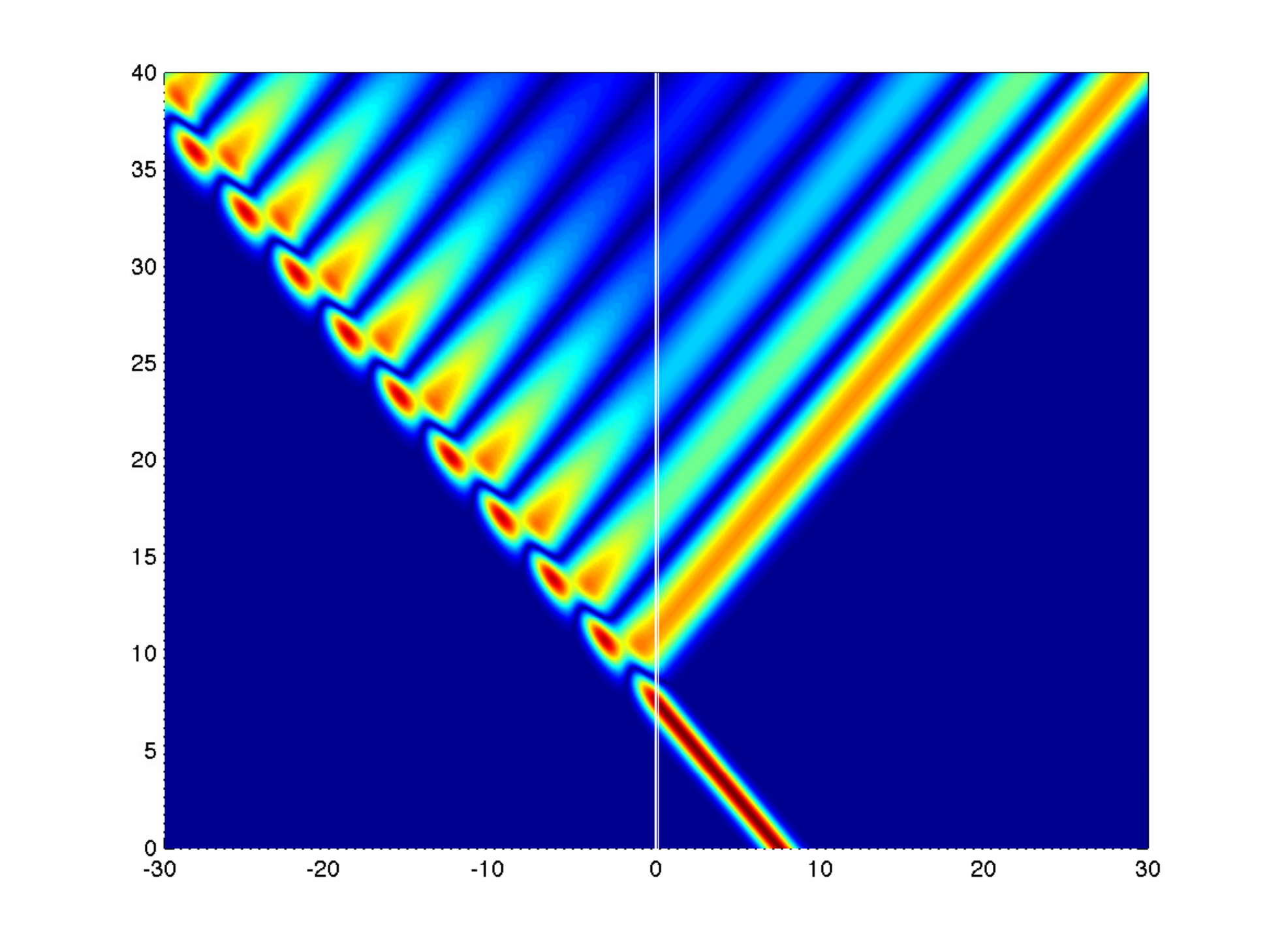} &
   \hspace{-0.4in}
   \includegraphics[width=.5\textwidth]{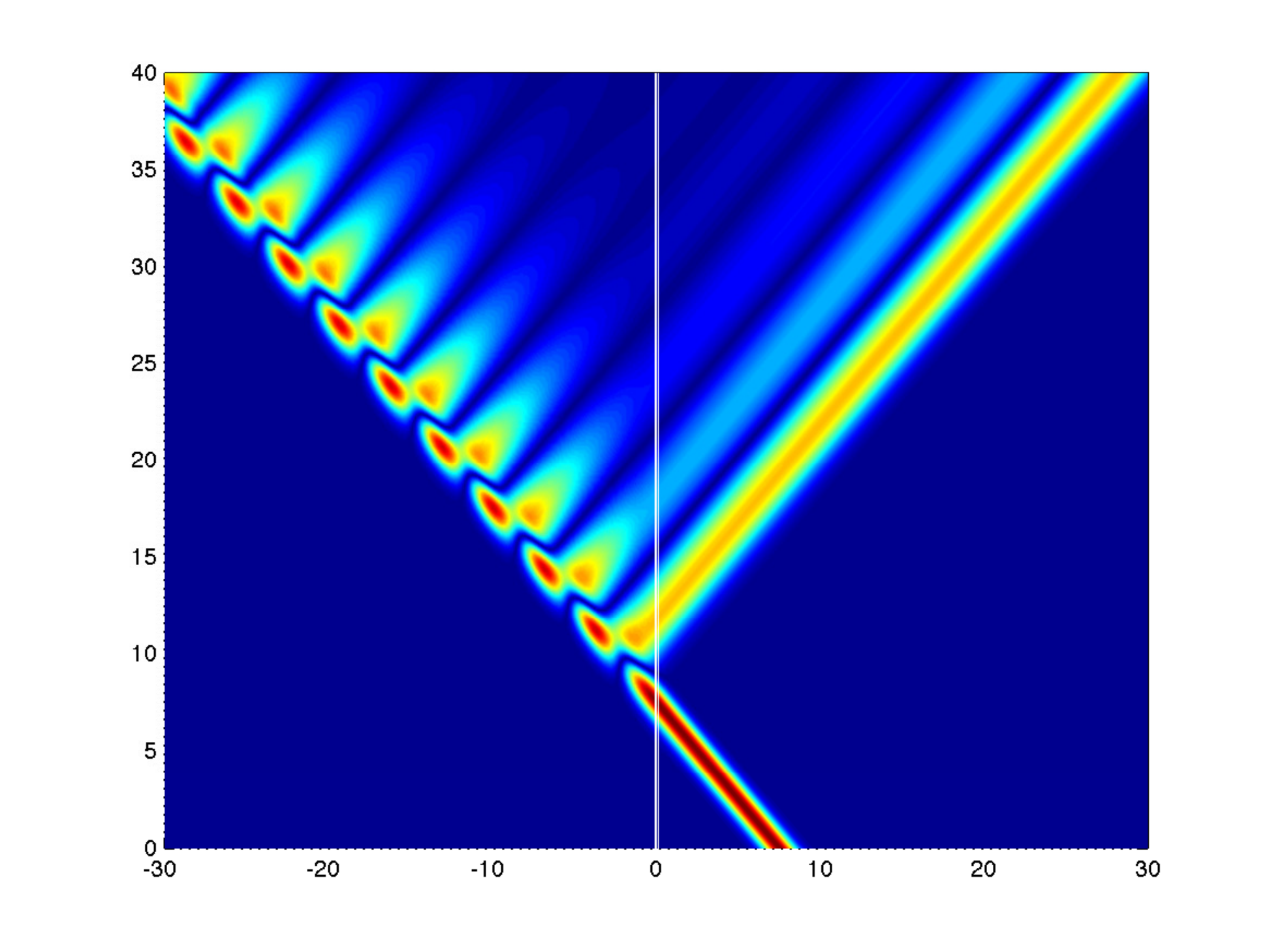}
  \end{tabular}
  \caption{$(x,t)$ plots of the solution 
amplitude computed with $\alpha =1$, $\beta =0$ and the four values of $L$~: $0$, $0.5$, $1$ and $2$. The vertical white line represents $x=0$.} \label{LregWidth}
\end{figure}

\begin{figure}[ht!]\noindent\makebox[\textwidth][c]{\begin{minipage}[t]{10cm}
\includegraphics[width=10cm,height=7.cm]{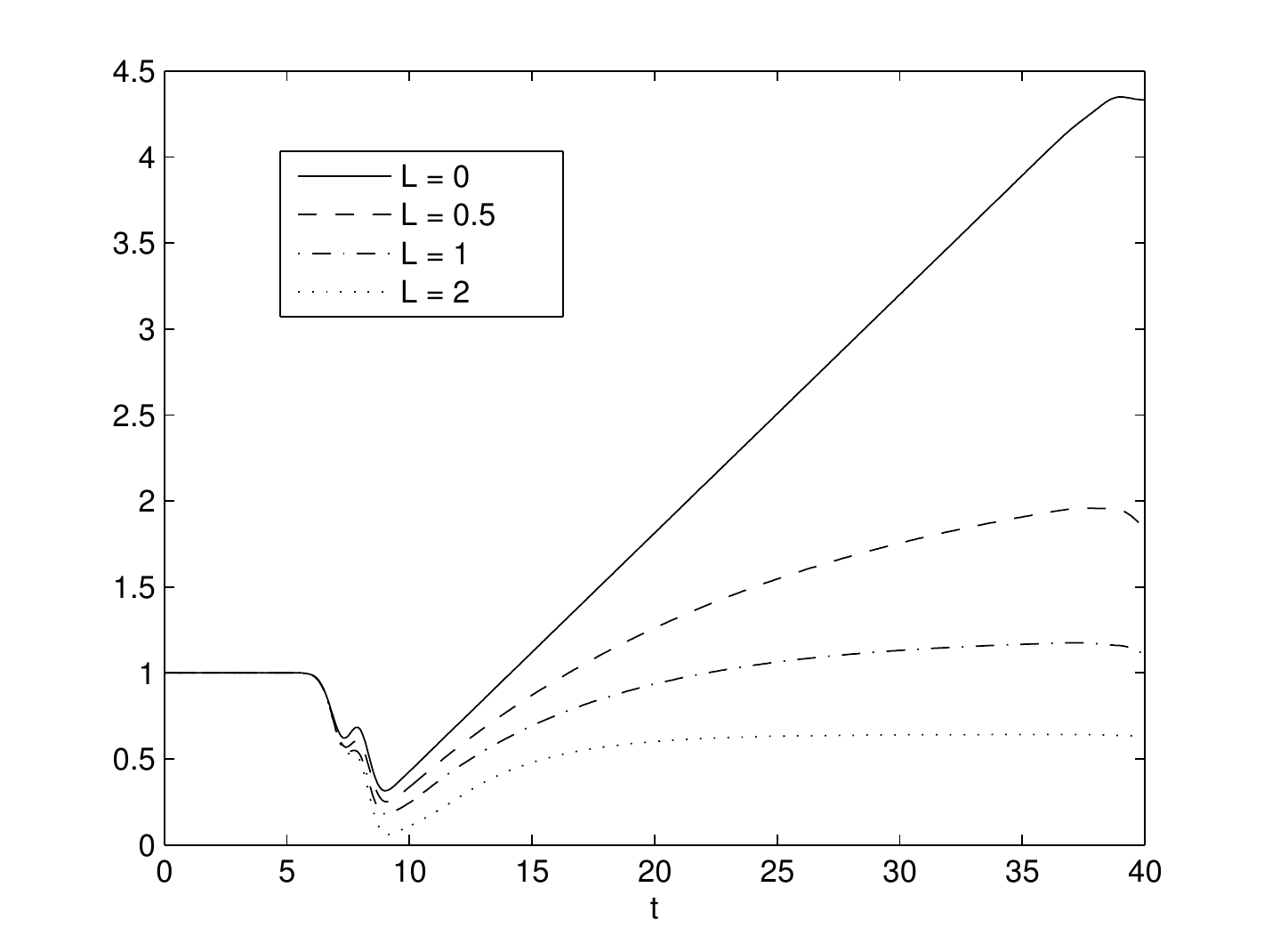}
\vspace*{-0.5cm}
\caption{Temporal profile of the energy
gain $G$ computed with different $L$~: $L=0$,
$L=0.5$, $L=1$ and $L=2$.}
\label{fig14}
\end{minipage}
}
\end{figure}

We first investigate the influence of the smoothing parameter $L$ on the
gain of energy, recalling that in the limit case $L=0$ the two potentials
are discontinuous if not identically zero. We only consider the case 
$\alpha \neq 0$ (there is no superradiance if $V\equiv 0$) and $\beta =0$ (i.e. $P\equiv 0$)~; we can then impose $\alpha = 1$ via a rescaling $t \mapsto a t$, $x \mapsto ax$. 
In Figure \ref{LregWidth}, we plot the $(x,t)$ profile of the absolute value of the real part of the solution computed on the spatial domain $[-30,30]$ until final time $T=40$. The value of $x_0$ is $7.5$~; numerically, this gives data that are compactly supported and whose support is contained in the positive real axis. The solution propagates freely towards the left until it enters the negative real axis.
Then, the effect of the smoothing of the potential on 
the solution is clearly observed. We compute for each simulation the gain
profile $G(t):=E_+ (\phi (t))/ E_+(\phi (0))$ that we plot in Figure 
\ref{fig14}. We find a linearly increasing profile for
the discontinuous case $L=0$, the gain is unbounded, hyperradiance
is observed. For increasing values of $L$ the gain becomes bounded and
the limit value turns out to be a decreasing function
of $L$. Superradiance is observed for $L\le 1$ and it is found that the amount of
superradiance depends on the smoothness parameter $L$. Hyperradiance is only observed for $L=0$.

Note that here, the computation of the gain has been made until
final time $T=40$. When considering simulations performed for larger
time, we would find (even for $L=0$)
gains that would become stationary due to the choice
of the spatial domain. Indeed, using \eqref{GainToyModel}, 
the energy is computed as an integral 
over the interval $[0,L]$. Since transparent 
conditions are used, waves go outside this domain and a part 
of the energy will not be taken into
account for the computation of
(\ref{EnergyComp}). This implies that when dealing with this kind of energy
evaluation, the choice of spatial bound of the computational domain
is drastically linked to the prescribed final time $T$ 
(we have to set $L\approx T$ to calculate a solution that remains
supported in the computational domain). Aiming to 
calculate the asymptotics for large times, it is not convenient
to measure the energy gain with $G$. 

To remedy this difficulty, we give an alternative definition of the gain using geometrical fluxes. We use the fact that for $x>0$ the toy model equation is exactly the one-dimensional Klein-Gordon equation
\[ \partial_t^2 \phi - \partial_x^2 \phi + \beta \phi = 0 \]
for which we have a conserved stress-energy tensor
$$
T_{ab} = \sum_{j=1}^2 \left( \partial_a \phi_j \partial_b \phi_j - \frac{1}{2} 
\big((\partial_t \phi_j)^2 - (\partial_x \phi_j)^2 -  
\beta \phi_j^2 \big) \eta_{ab} \right)
$$
with $\eta = \d t^2 - \d x^2$, $\phi_1 = \Re \phi$ and 
$\phi_2 = \Im \phi$. This gives us an energy current $J^a = T_0^a$ 
whose fluxes across $\{t \} \times [0,+\infty [ $ and $[0,t] \times 
\{R \}$ are given by
\begin{eqnarray*}
{\cal F}_{\{t \} \times [0,+\infty [} &=& \frac{1}{2} \int_{\R^+} ( \vert \partial_t \phi (t,x) \vert^2 + \vert \partial_x \phi (t,x) \vert^2
+\beta \vert \phi(t,x)\vert^2 ) \d x \, , \\
{\cal F}_{[0,t] \times \{R \}} &=& -\int_0^t ( \partial_t \phi_1 (\tau ,R) \partial_x \phi_1 (\tau ,R) + \partial_t \phi_2 (\tau ,R) \partial_x \phi_2 (\tau ,R) ) \d \tau \, .
\end{eqnarray*}
We use the alternative definition of the gain
\begin{equation} \label{GainFlux}
\tilde G_{R}(t) := \frac{1}{{\cal F}_{\{0\} \times [0,+\infty [}} 
{\cal F}_{[0,t] \times \{R\}}
\end{equation}
for some $R\gg1$ large enough so that the support of the initial data is contained in the interval $[0,R]$. The point of using the outgoing energy flux to calculate the gain is that we can choose smaller spatial intervals on which to calculate, taking advantage of our transparent boundary conditions, as will be seen in our numerical results.

\begin{center}
\begin{figure}[ht!]
\begin{minipage}{7.4cm}
\includegraphics[width=7.4cm,height=6.4cm]{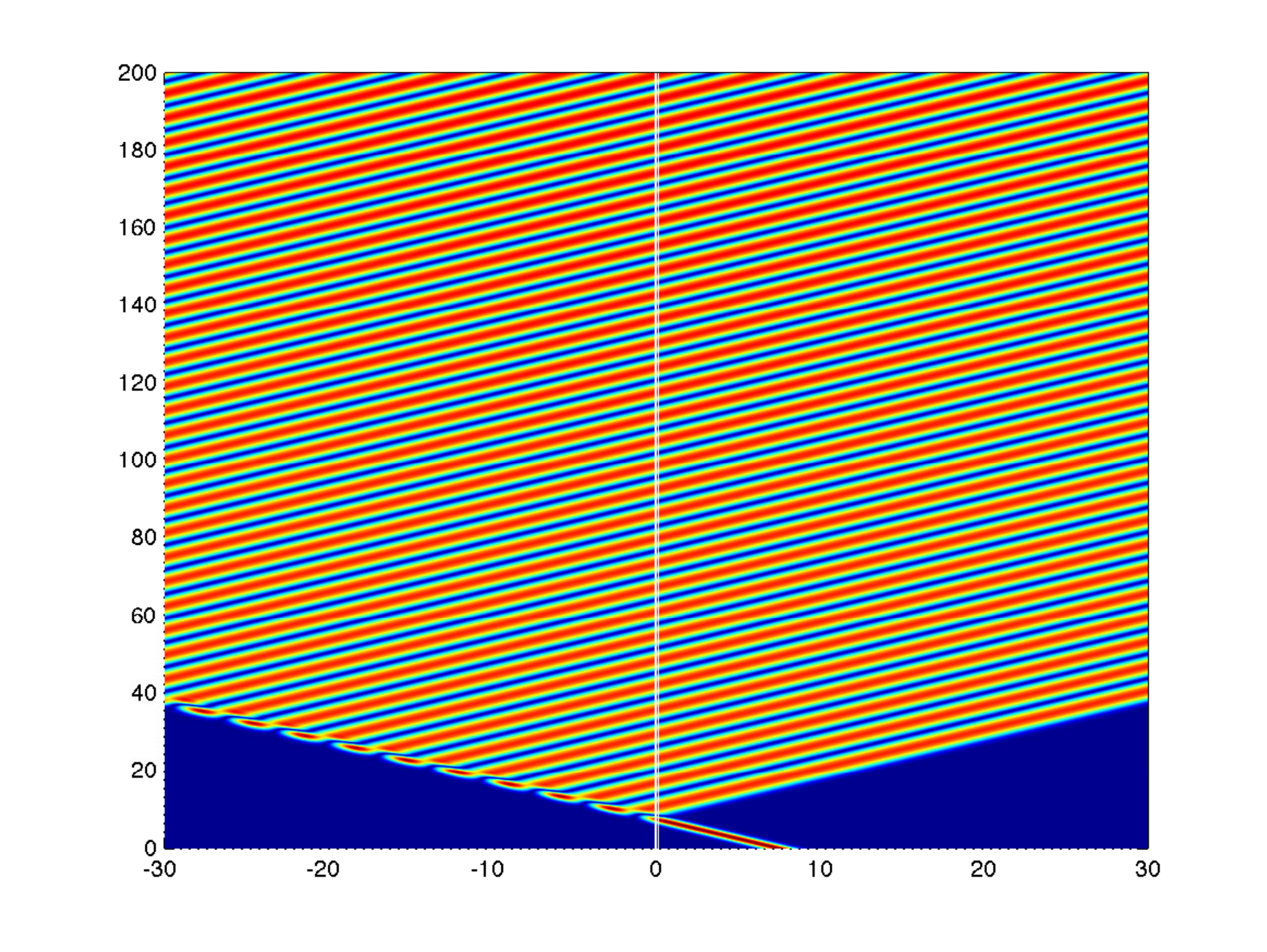}
\vspace*{-0.9cm}
\caption{$(x,t)$ plot of the solution 
amplitude, hyperradiant case $\alpha =1$, $\beta =0$, $L=0$.}
\label{fig15}
\end{minipage}
\hspace*{0.2cm}
\begin{minipage}{8.4cm}
\includegraphics[width=8.4cm,height=6.4cm]{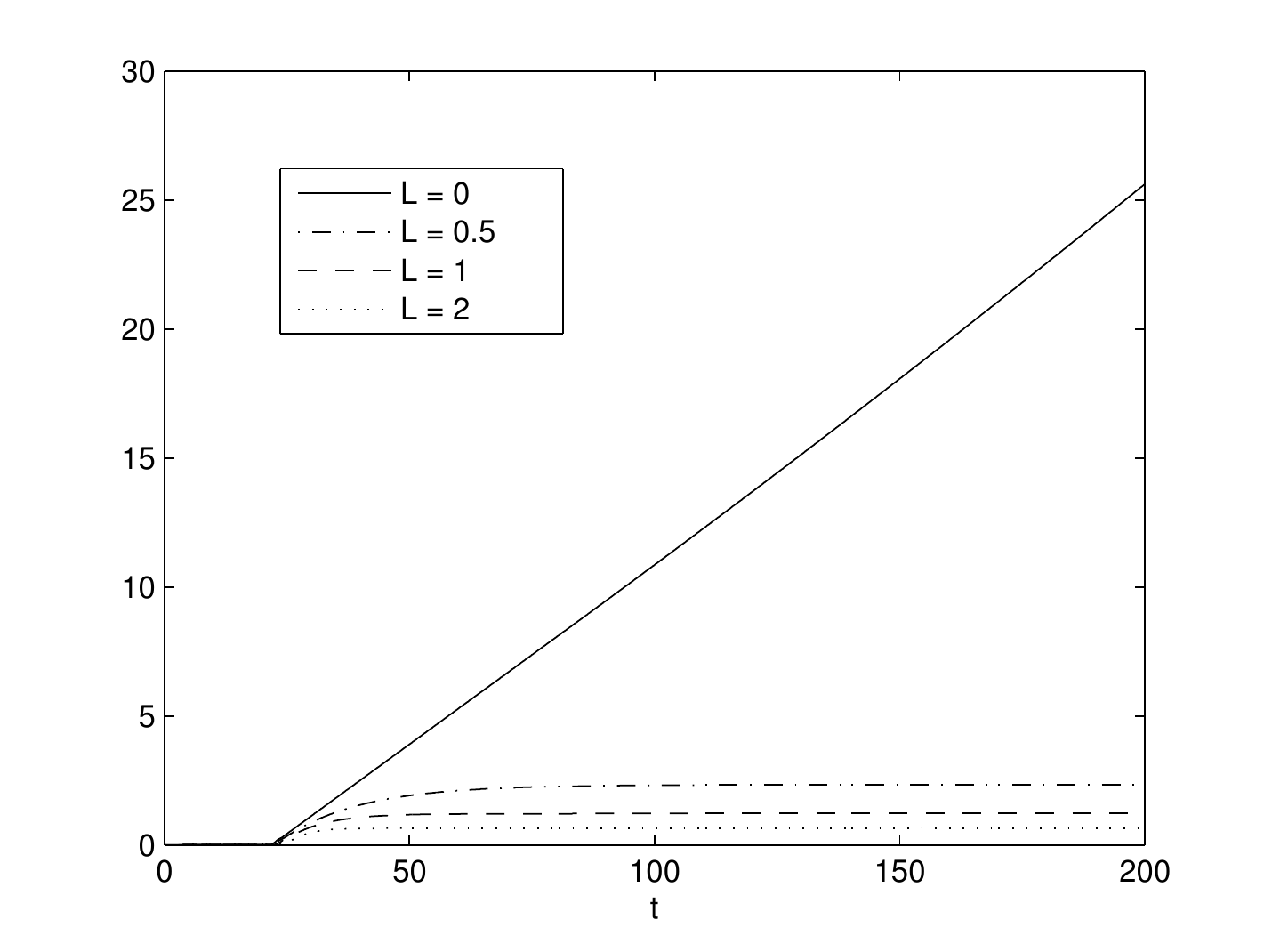}
\vspace*{-0.9cm}
\caption{Plot of the gain computed with \eqref{GainFlux} in the case $\alpha =1$, $\beta =0$ for
$L=0$, $L=0.5$, $L=1$ and $L=2$.}
\label{fig16}
\end{minipage}
\end{figure}
\end{center}

For the same experiment, we now compute the gain using 
(\ref{GainFlux}) until larger times, for the same values of
$L$ as before.
As can be seen in Figure \ref{fig15} where the solution amplitude
is plotted, the solution has spread outside the domain and it 
becomes impossible to calculate the energy gain in terms of $G$. We then notice in Figure  \ref{fig16} that 
the gain now becomes unbounded even if the amount of energy 
contained inside the computational domain is constant for $t\ge 40$. 
This clearly suggests that the gain measurement in terms of flux across a 
timelike hypersurface is preferable for long-time simulations. This
is a consequence of the use of well-adapted boundary conditions
at the segment extremities.

\subsection{Numerical results for the Reissner-Nordstr\o m case}

When trying to find numerically solutions of \eqref{CKGRNphi} with a superradiant behaviour, we must be careful to work with solutions for which the meaning of the energy gain is unambiguous. There is no particular precaution to take with the outgoing flux, which has a clear-cut geometrical definition, but it is important to specialize to subspaces of initial data on which the energy is positive definite. We consider two types of such data.
\begin{enumerate}
\item {\it Incoming wave packets.} The data are located far outside the effective ergosphere, which requires to chose physical parameters such that the effective ergosphere does not extend to infinity. This is guaranteed as soon as we choose the mass of the particle to be non zero. The data $\phi_0 = \phi \vert_{t=0}$ and $\phi_1 = \partial_t \phi \vert_{t=0}$ are chosen as follows~:
\begin{eqnarray}
\phi_0 (r_*) &=& e^{i \omega r_*/\lambda} e^{-(r_*-r_*^0)^2/\lambda^2} \, ,\\
\phi_1 (r_*) &=& \partial_{r_*} \phi_0 (r_*) \, .
\end{eqnarray}
For Equation \eqref{CKGRNphi}, the associated solution propagates dominantly towards the left at speed $1$, with a little dispersion, until it reaches the effective ergosphere.
\item {\it Flares.} We take
\begin{eqnarray}
\phi_0 (r_*) &= & 0  \, ,\\
\phi_1 (r_*) &=& e^{-(r_*-r_*^0)^2/\lambda^2} \, .
\end{eqnarray}
The energy is positive definite on such data since it is merely the $L^2$ norm of $\phi_1$. This is true independently of the location of the centre of the Gaussian. This allows us to consider cases where the effective ergosphere covers the whole domain of outer communication. We will not do so in this paper however, since we wish to compare the gains we obtain with flares and incoming wave packets. We shall therefore keep the same values of the physical parameters, and in particular a non-zero mass for the field, in our simulations with both types of data.
\end{enumerate}
In both cases we measure an outgoing energy flux to the right of the support of the initial data.

\subsubsection{Incoming wave packets}

\begin{figure}[ht!]
\centering
  \begin{tabular}{@{}cc@{}}
   \hspace{-0.3in}
   \includegraphics[width=.55\textwidth]{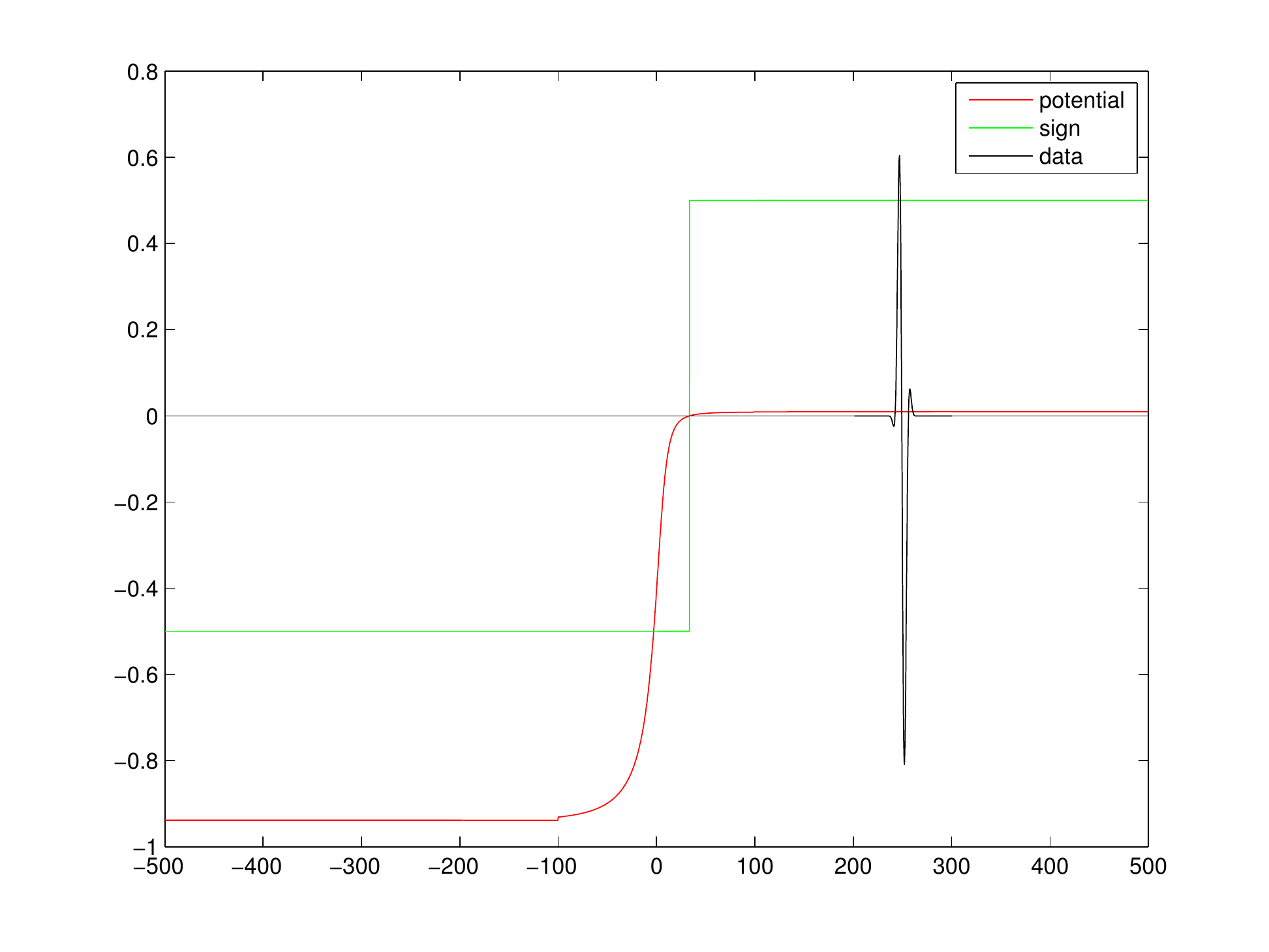} &
   \hspace{-0.4in}
   \includegraphics[width=.55\textwidth]{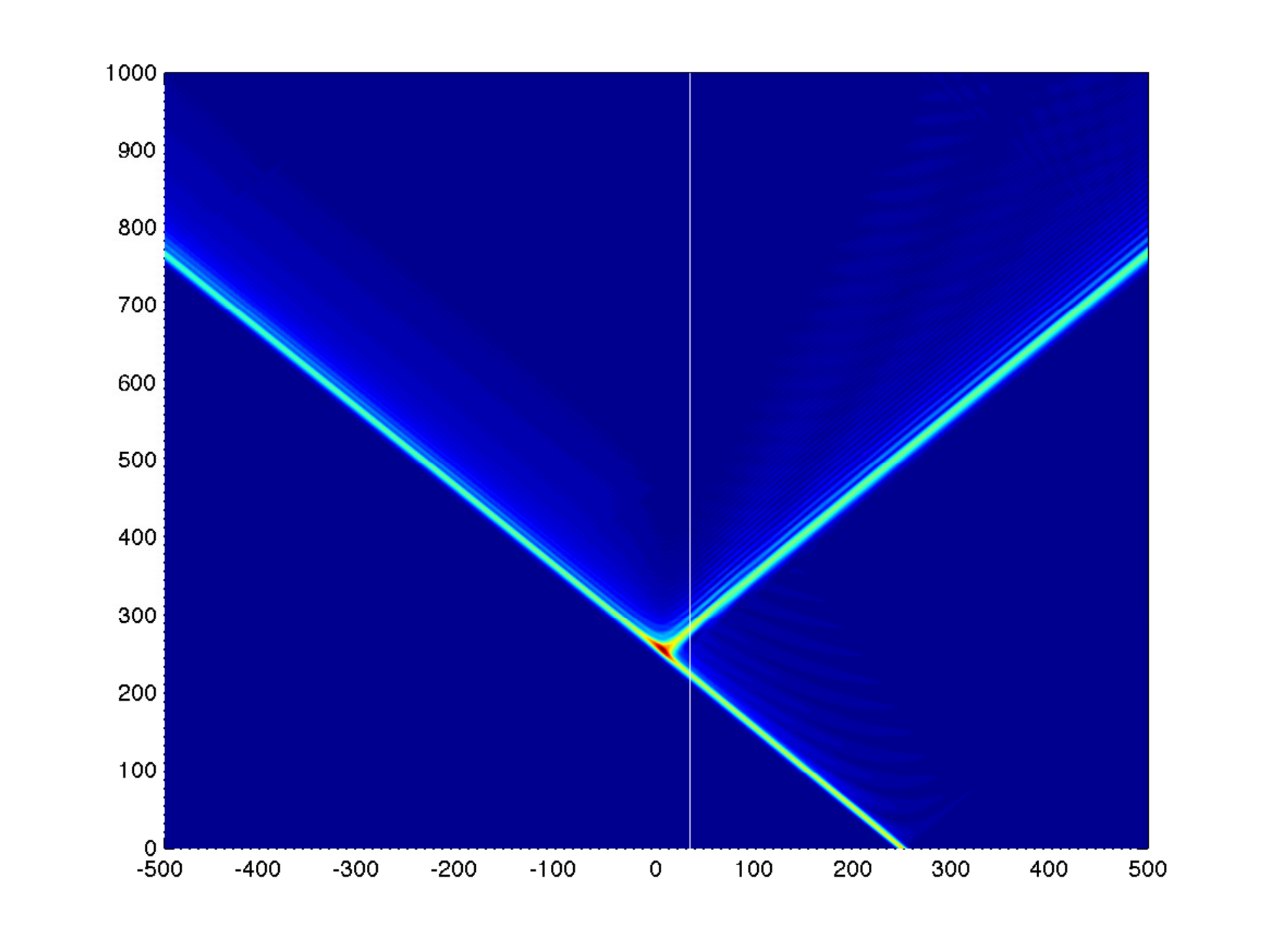}
  \end{tabular}
  \caption{For $\omega = 2.3$, the potential, its sign and the real part of the initial data $\phi_0$ for our choice of physical parameters, then the evolution of the field. The frontier of the effective ergosphere, where the potential changes sign, is located at $r_* \simeq 33.67$ and is represented as a vertical white line on the second diagram.} \label{RNMMaxGain}
\end{figure}
\begin{center}
\begin{figure}[ht!]
\centering\epsfig{figure=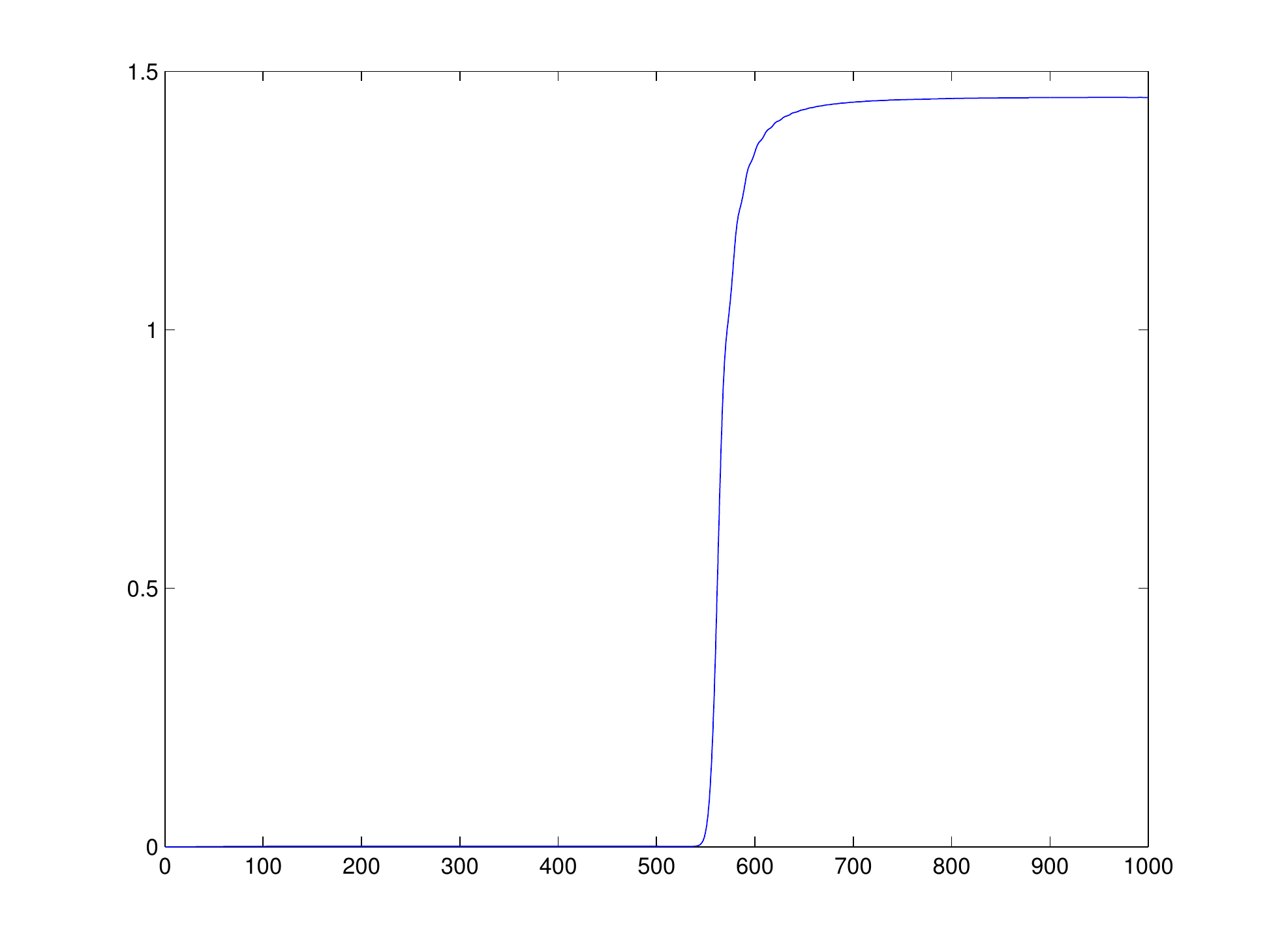,width=10cm,height=7cm}
\caption{Time evolution of the associated energy gain, stabilizing at ${\cal G}_\infty \simeq 1.449$.}\label{RNMER}
\end{figure}
\end{center}
The toy model showed us that superradiance is driven by the steepness of the transition of the potential between its two limit values at $-\infty$ and $+\infty$. We choose physical parameters that make the potential
\[ F(r) \left( \frac{l(l+1)}{r^2} + m^2 + \frac{F'(r)}{r} \right) - \frac{q^2 Q^2}{r^2}\]
look like a narrow regularization of a step function. We give here an example of simulation corresponding to the following values~:
\begin{equation} \label{PhysicalParameters}
M = 2.001\, , ~ Q=2 \, ,~ m=0.1 \, ,~ q =1 \, ,~ l=0 \mbox{ (spherical symmetry)} \, .
\end{equation}
The numerical simulation is performed on the interval $[-500 , 500]$ in the variable $r_*$, with data centered at $r_*^0 = 250$ and a scaling parameter $\lambda =5$. The outgoing energy flux is measured at $r_* = 300$. The highest gain is reached for the frequency $\omega = 2.3$. The potential, its sign and the real part of the initial data $\phi_0$ as well as the evolution of the field are plotted in Figure \ref{RNMMaxGain}. The stabilization of the energy gain is shown in Figure \ref{RNMER}. No hyperradiance occurs, the gain stabilizes to a finite value. Nevertheless, superradiance is observed.

\begin{figure}[htb!]
\centering
  \begin{tabular}{@{}cc@{}}
   \hspace{-0.3in}
   \includegraphics[width=.55\textwidth]{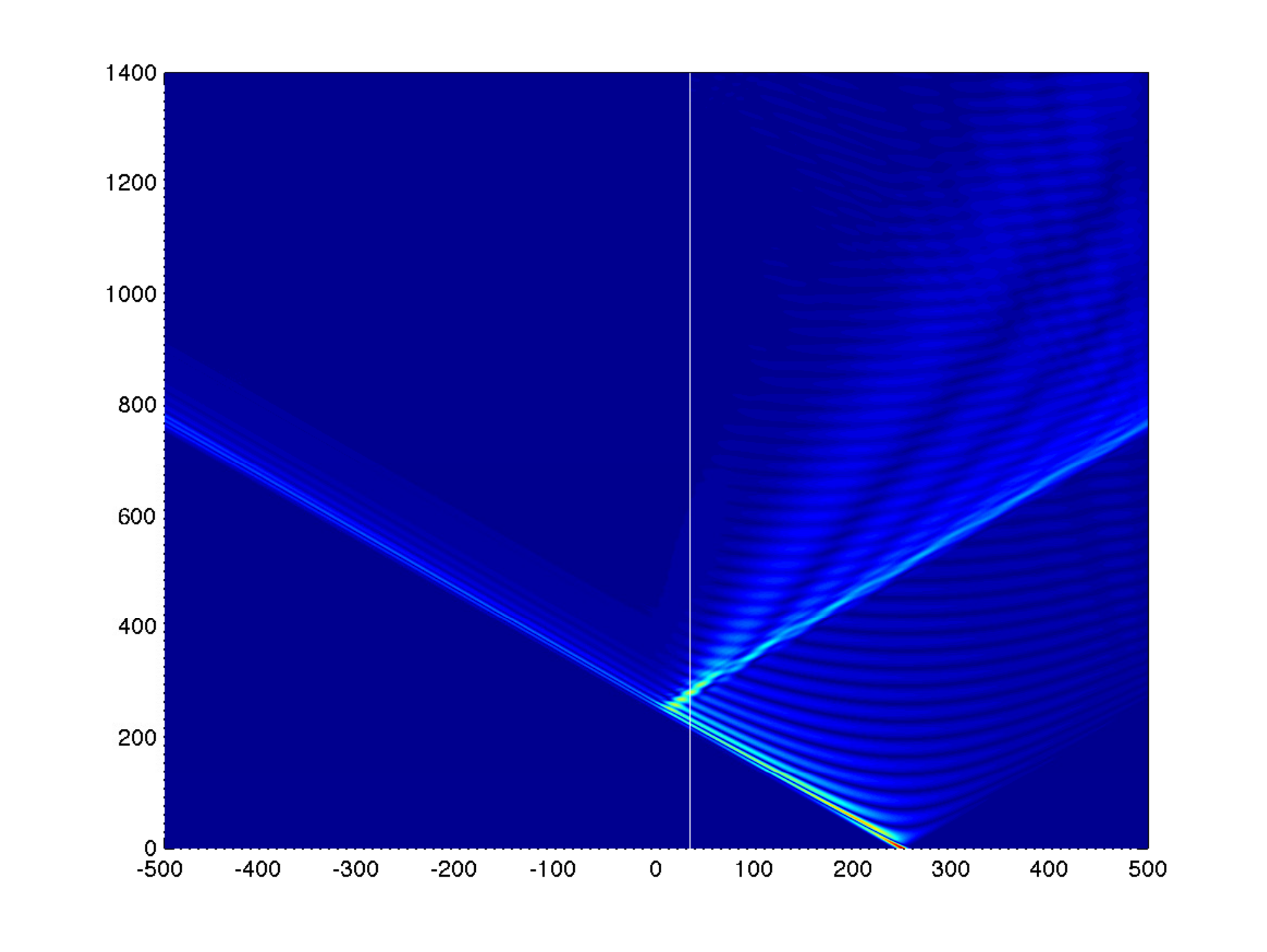} &
   \hspace{-0.4in}
   \includegraphics[width=.55\textwidth]{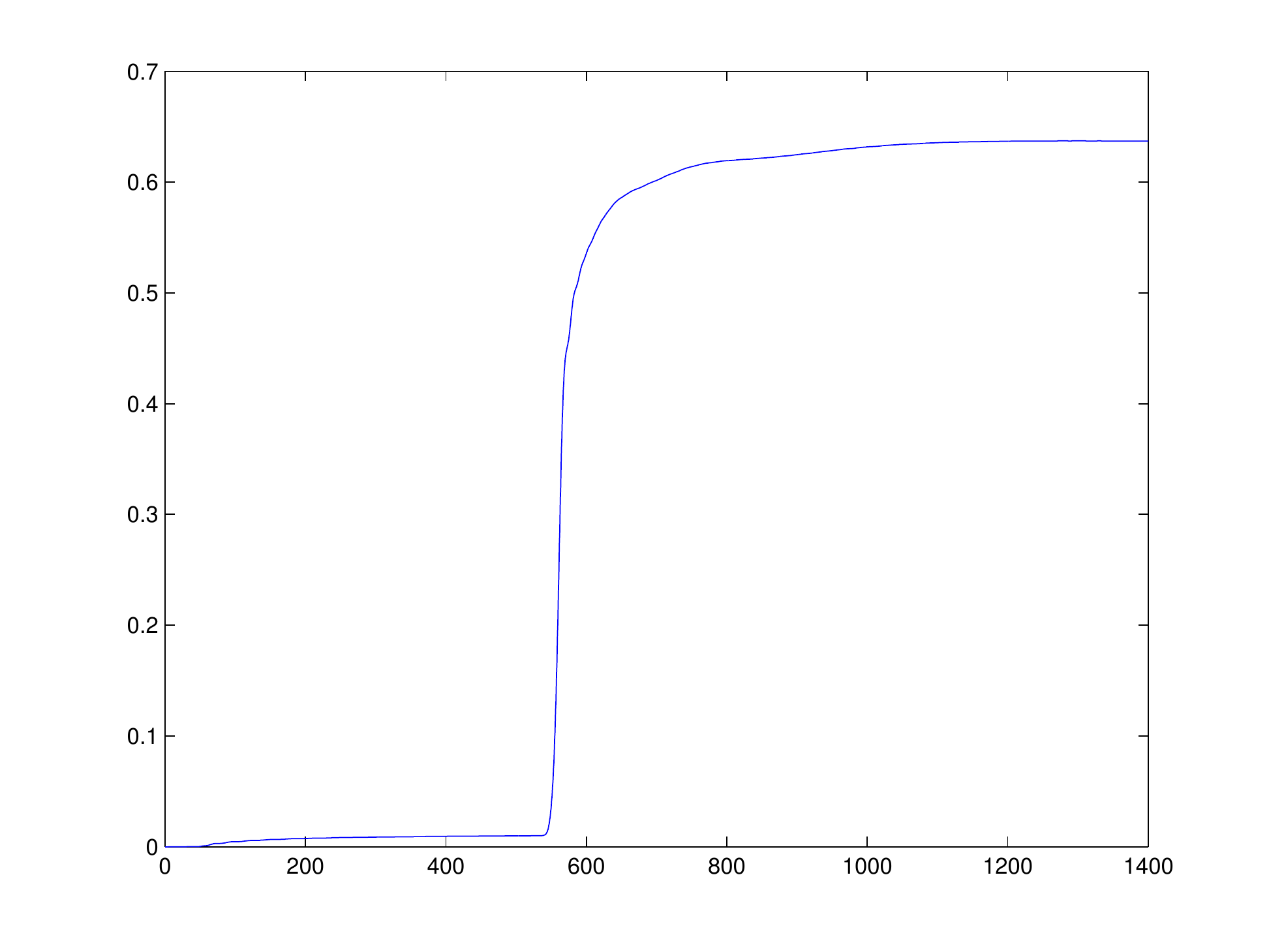} \\
   \hspace{-0.3in}
   \includegraphics[width=.55\textwidth]{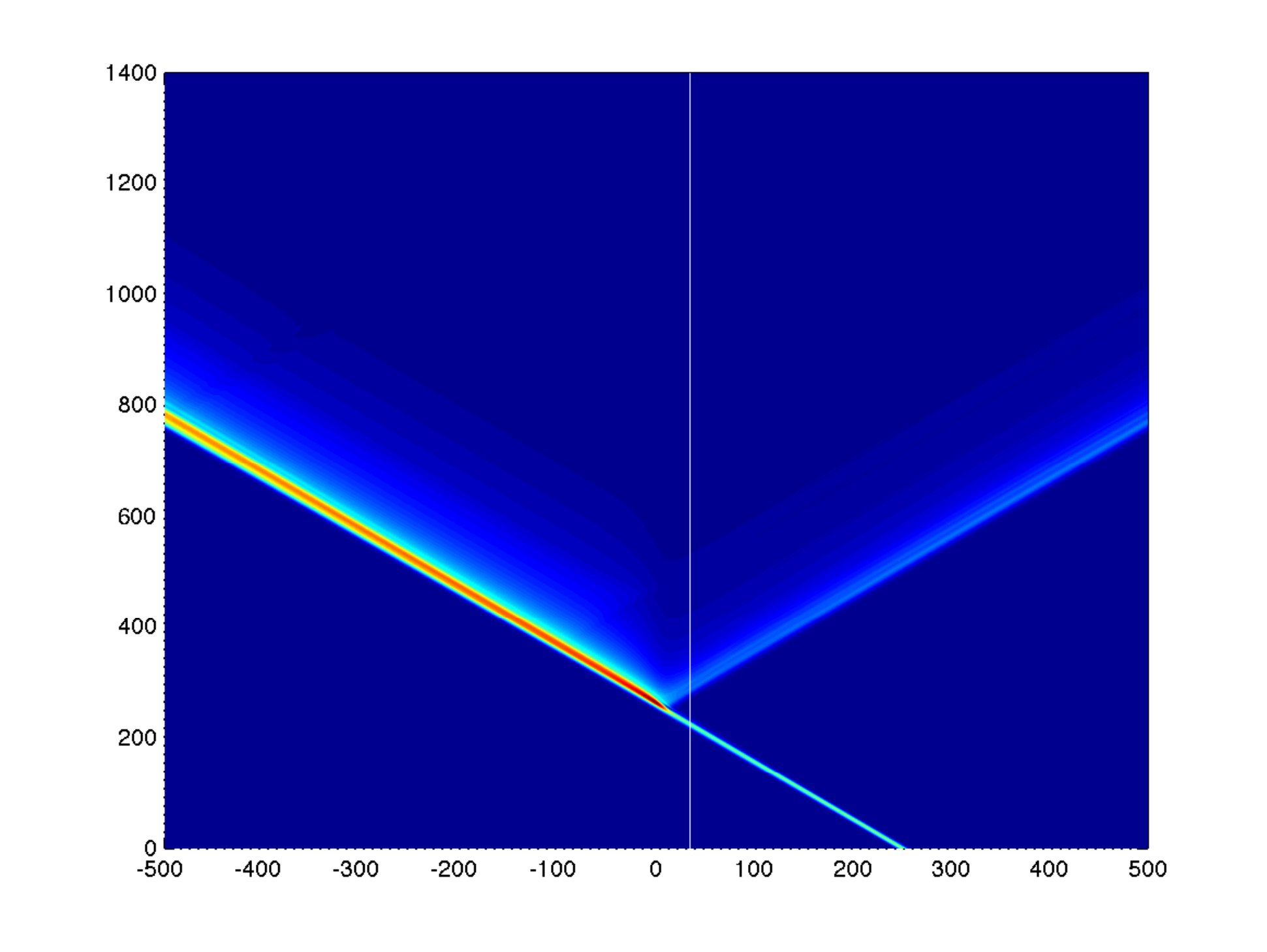} &
   \hspace{-0.4in}
   \includegraphics[width=.55\textwidth]{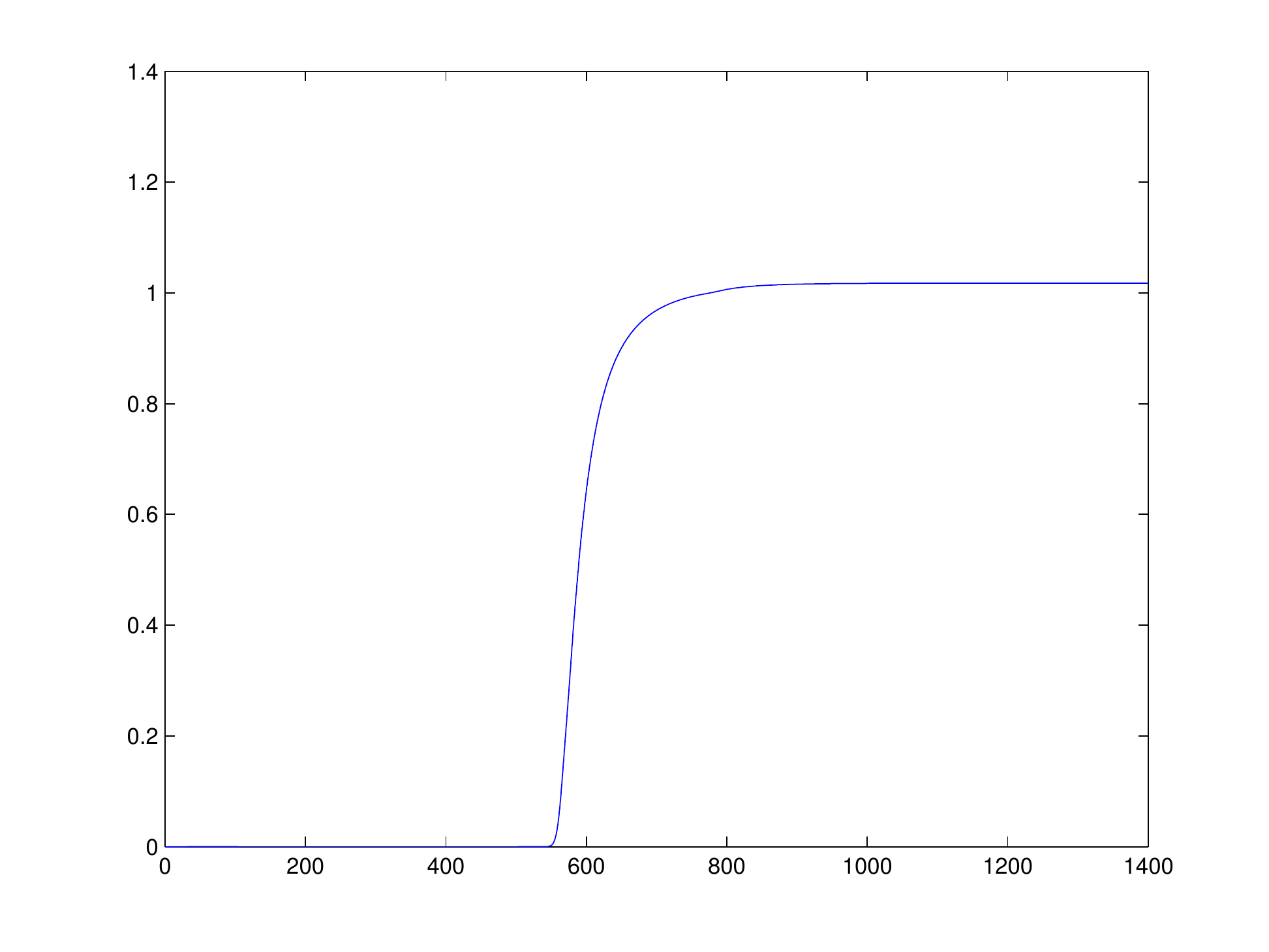} \\
   \hspace{-0.3in}
   \includegraphics[width=.55\textwidth]{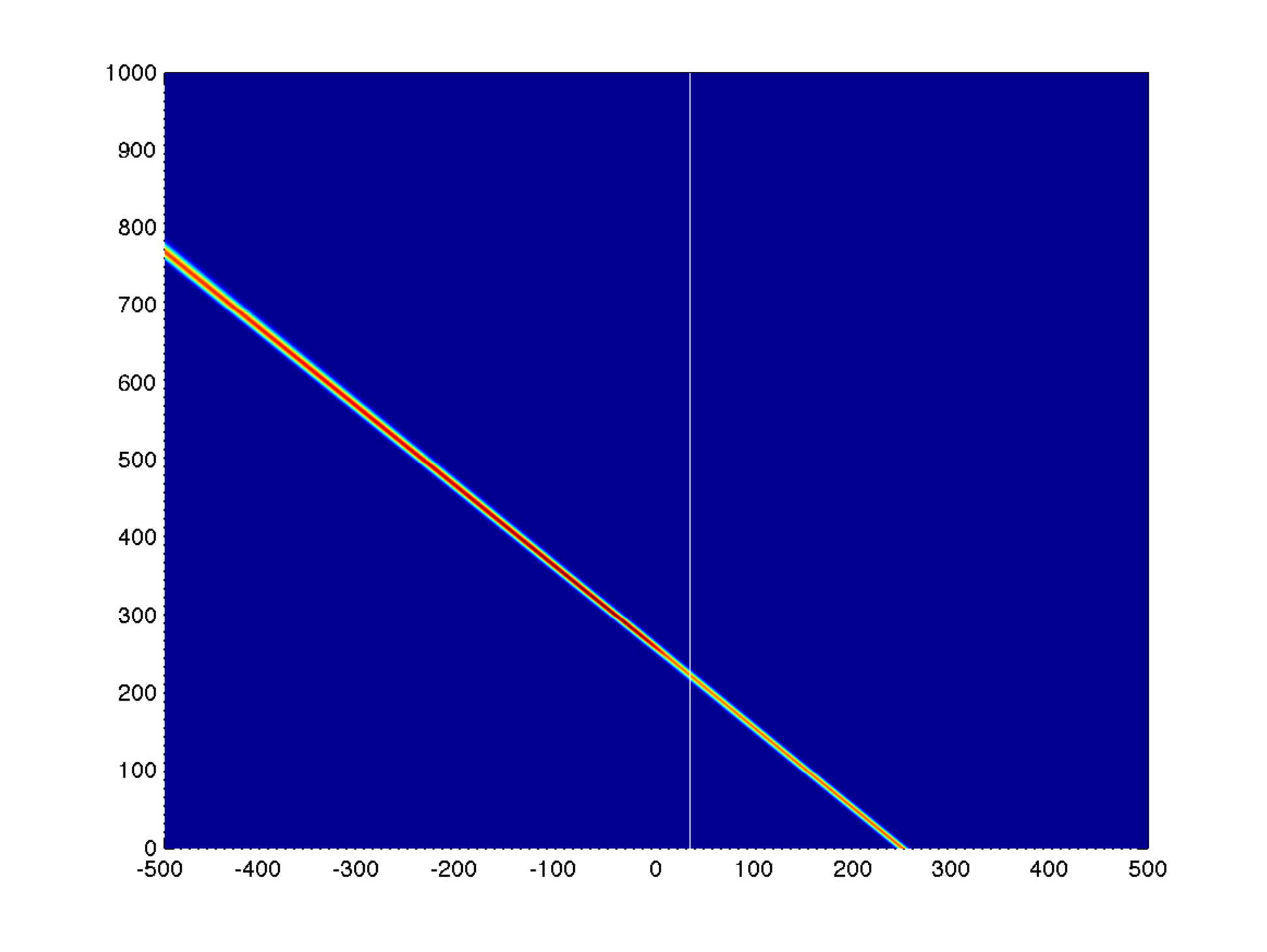} &
   \hspace{-0.4in}
   \includegraphics[width=.55\textwidth]{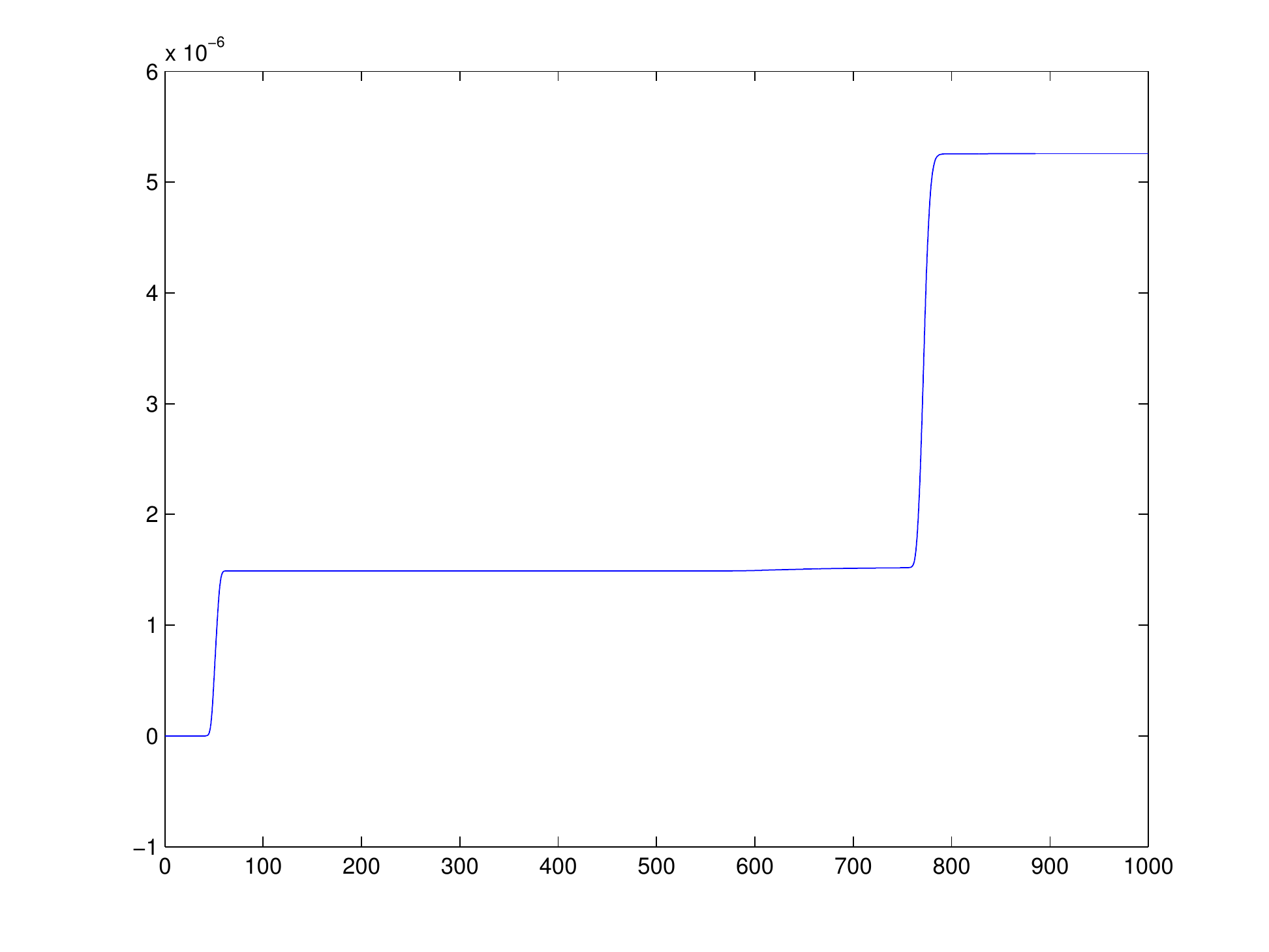}
  \end{tabular}
  \caption{Absolute value of the real part of the solution and associated energy gain for $\omega = 0$, $\omega=4$ and $\omega =10$.} \label{Missile}
\end{figure}

When we increase the frequency from the value above, the asymptotic gain decreases, the behaviour ceases to be superradiant from a value of $\omega$ a little above $4$, then the asymptotic gain rapidly becomes negligeable and the wave packet follows the incoming radial null geodesics ever more closely. Note that for higher frequencies, the gain stabilizes faster, which reflects the fact that the wave packet undergoes less dispersion as it propagates. Three examples are shown in Figure \ref{Missile}. This behaviour is consistent with the expected dynamics since superradiance is known to be a low-energy phenomenon.

\subsubsection{Flares}

\begin{figure}[htb!]
\centering
  \begin{tabular}{@{}cc@{}}
   \hspace{-0.4in}
   \includegraphics[width=.55\textwidth,height=7cm]{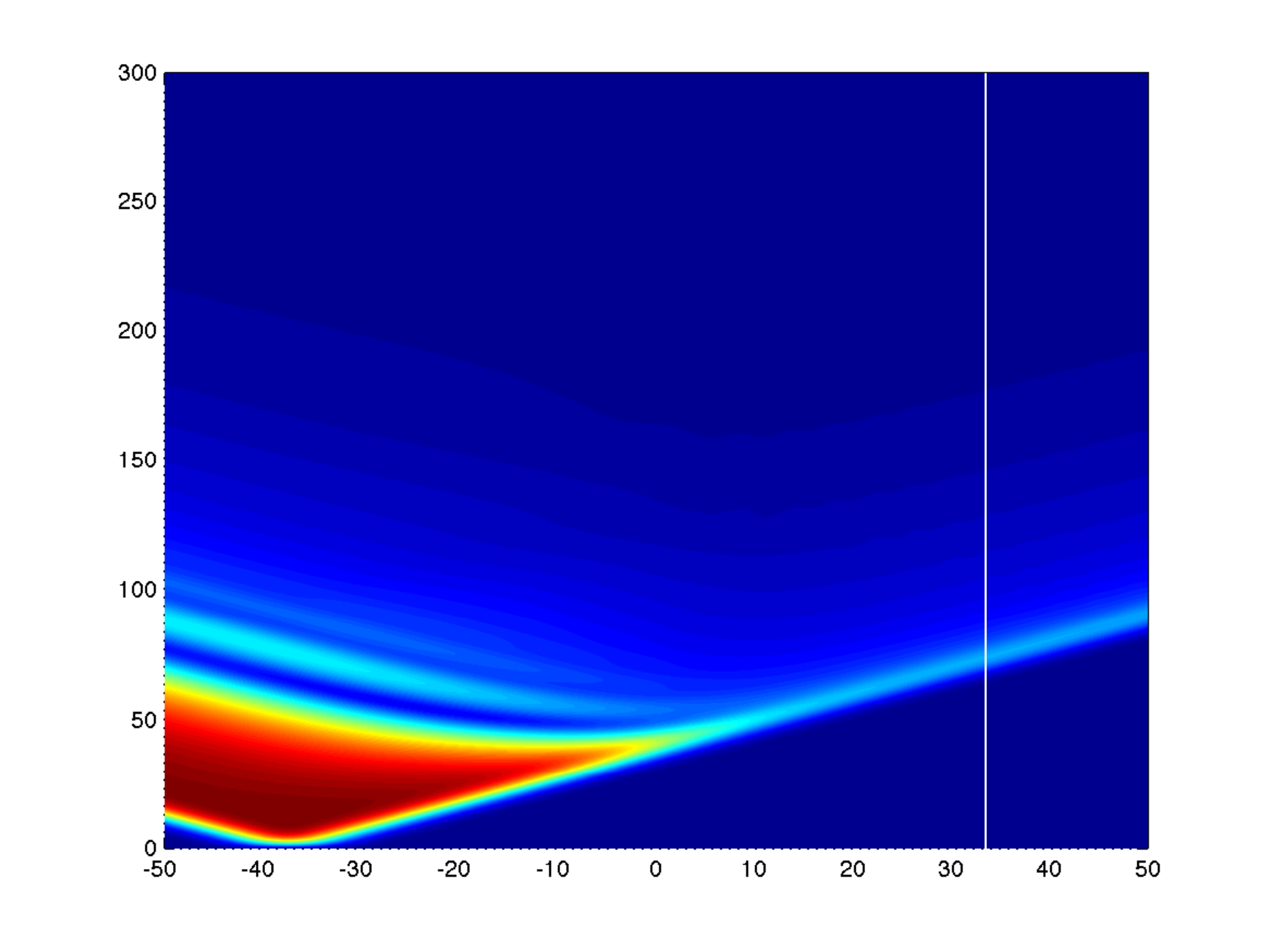} &
   \hspace{-0.4in}
   \includegraphics[width=.55\textwidth,height=7cm]{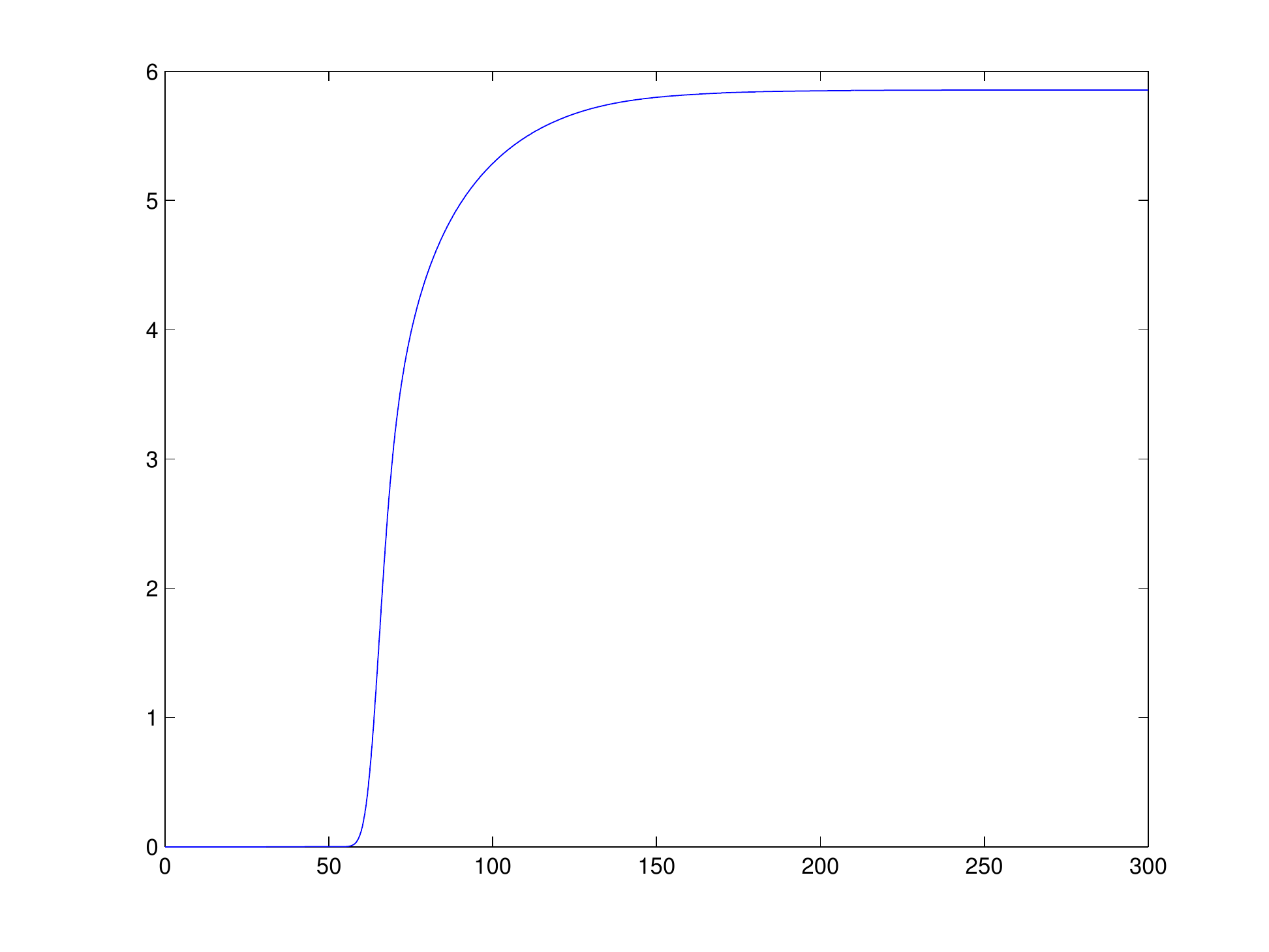}
  \end{tabular}
  \caption{An example of a flare-type initial data yielding a superradiant behaviour. The evolution of the absolute value of the real part of the solution is shown as well as the associated energy gain.} \label{Flare}
\end{figure}

Flare-type initial data, for the same physical parameters as before, also exhibit a superradiant behaviour but give a much larger gain. The evolution of the field is simulated on the interval $[-50,50]$ in the variable $r_*$, with data centered at $r_*^0 = -37.5$ and scaling factor $\lambda = 5$. The outgoing energy flux is measured at $r_*=1$. The values of the physical parameters are given by \eqref{PhysicalParameters}. Figure \ref{Flare} shows the evolution of the field and stabilization of the energy gain for this choice of parameters. We see that the gain is much more important than in the case of the incoming wave packet. Again, we observe no hyperradiant behaviour.

\subsubsection{High energy behaviour}

\begin{figure}[htb]
\centering
  \begin{tabular}{@{}ccc@{}}
   \hspace{-0.2in}
   \includegraphics[width=.35\textwidth,height=7cm]{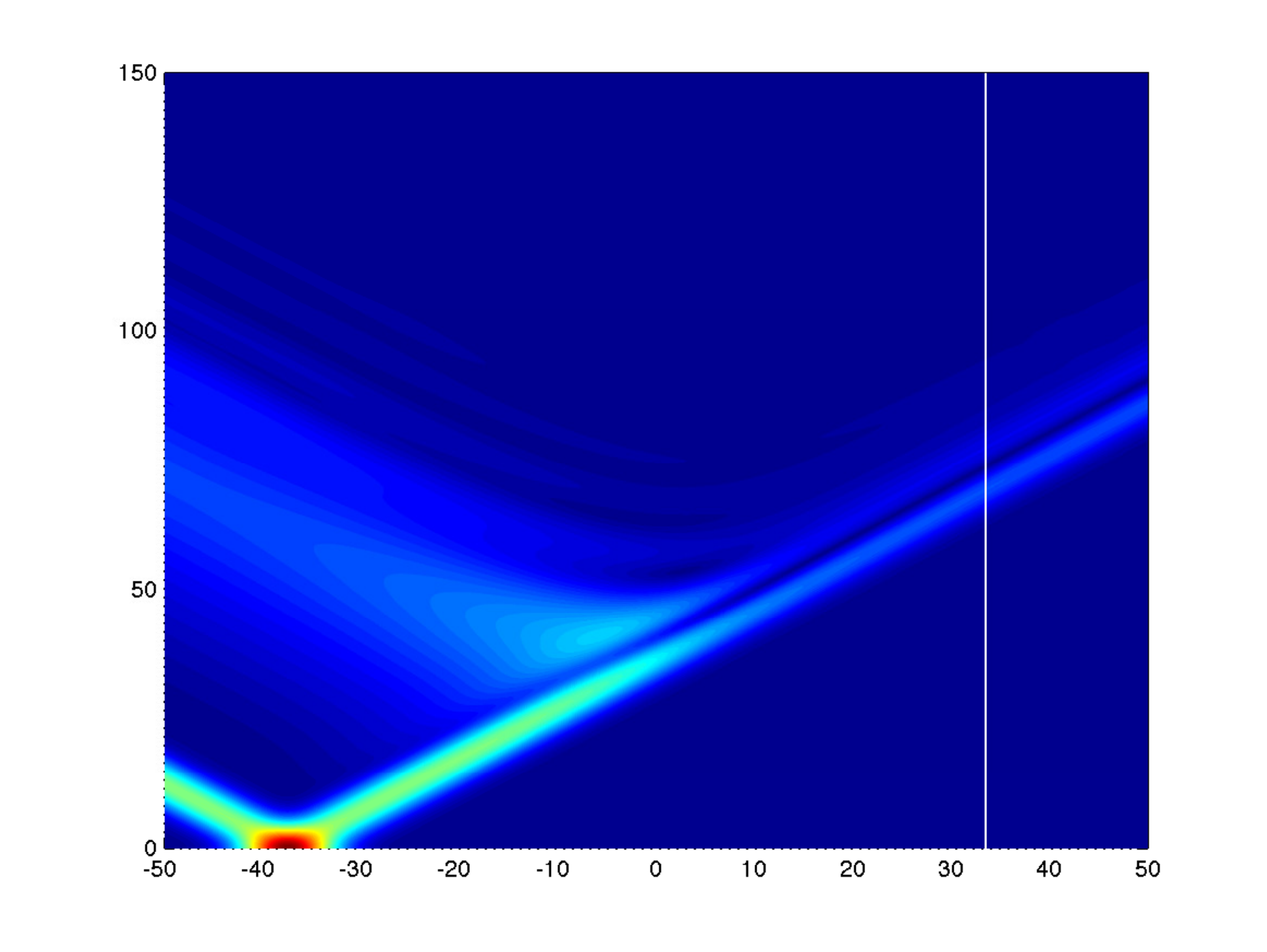} &
   \hspace{-0.4in}
   \includegraphics[width=.35\textwidth,height=7cm]{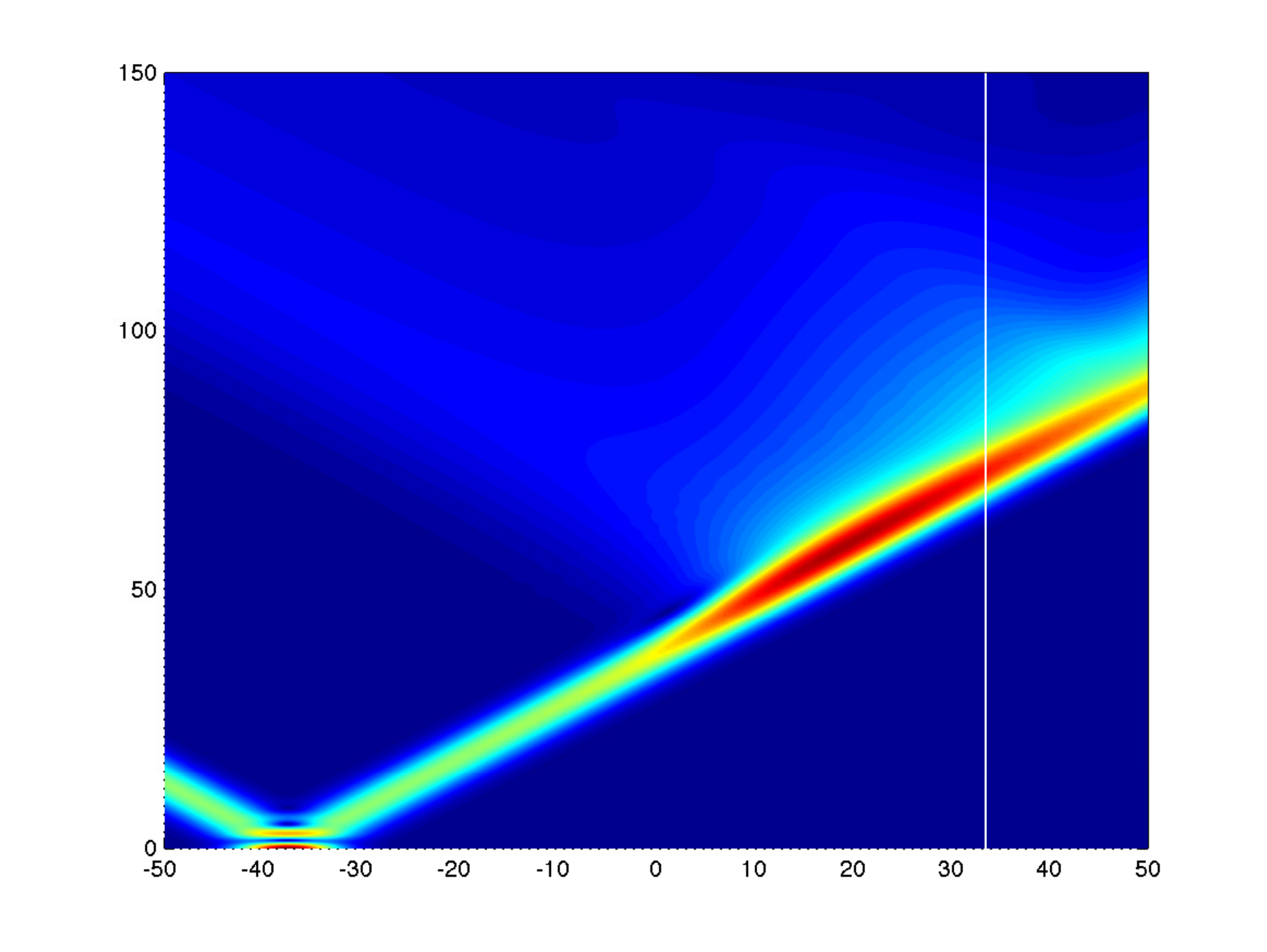} &
   \hspace{-0.4in}
   \includegraphics[width=.35\textwidth,height=7cm]{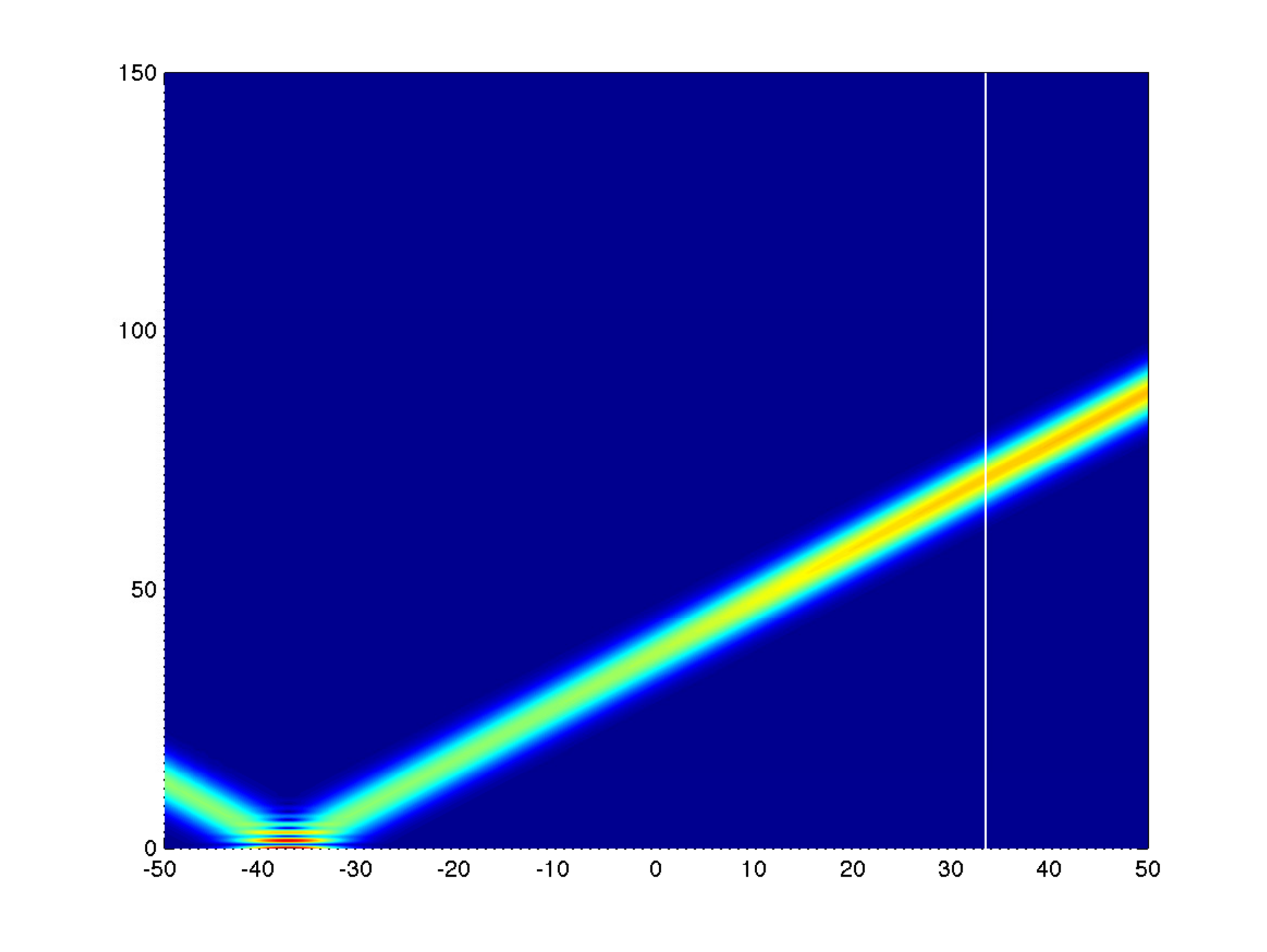}
  \end{tabular}
  \caption{Absolute value of the real part of the solution for $\omega =0$, $\omega=5$ and $\omega =10$.} \label{RHNE1}
\end{figure}

\begin{figure}[htb!]
\centering
  \begin{tabular}{@{}ccc@{}}
   \hspace{-0.2in}
   \includegraphics[width=.35\textwidth,height=7cm]{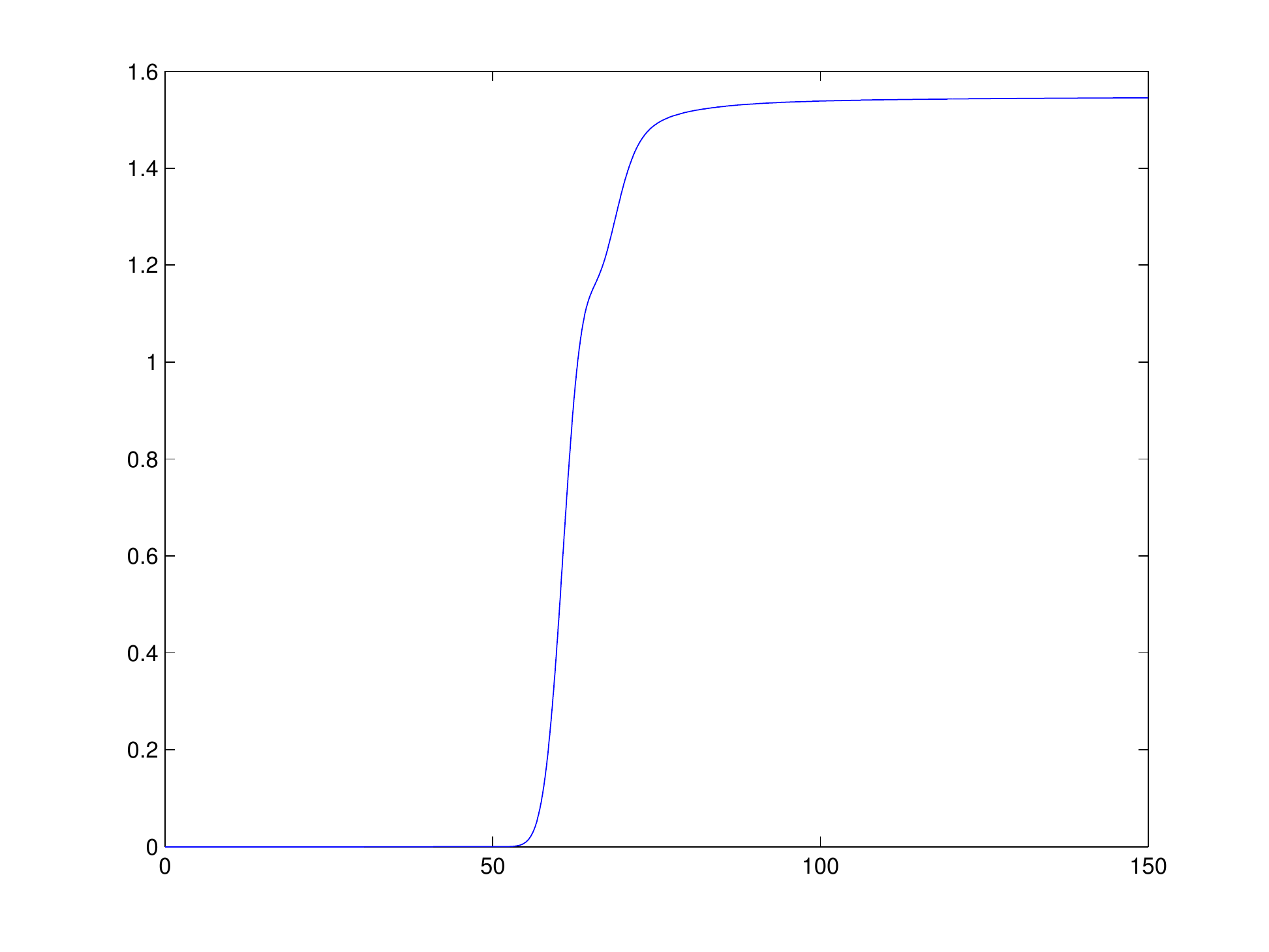} &
   \hspace{-0.4in}
   \includegraphics[width=.35\textwidth,height=7cm]{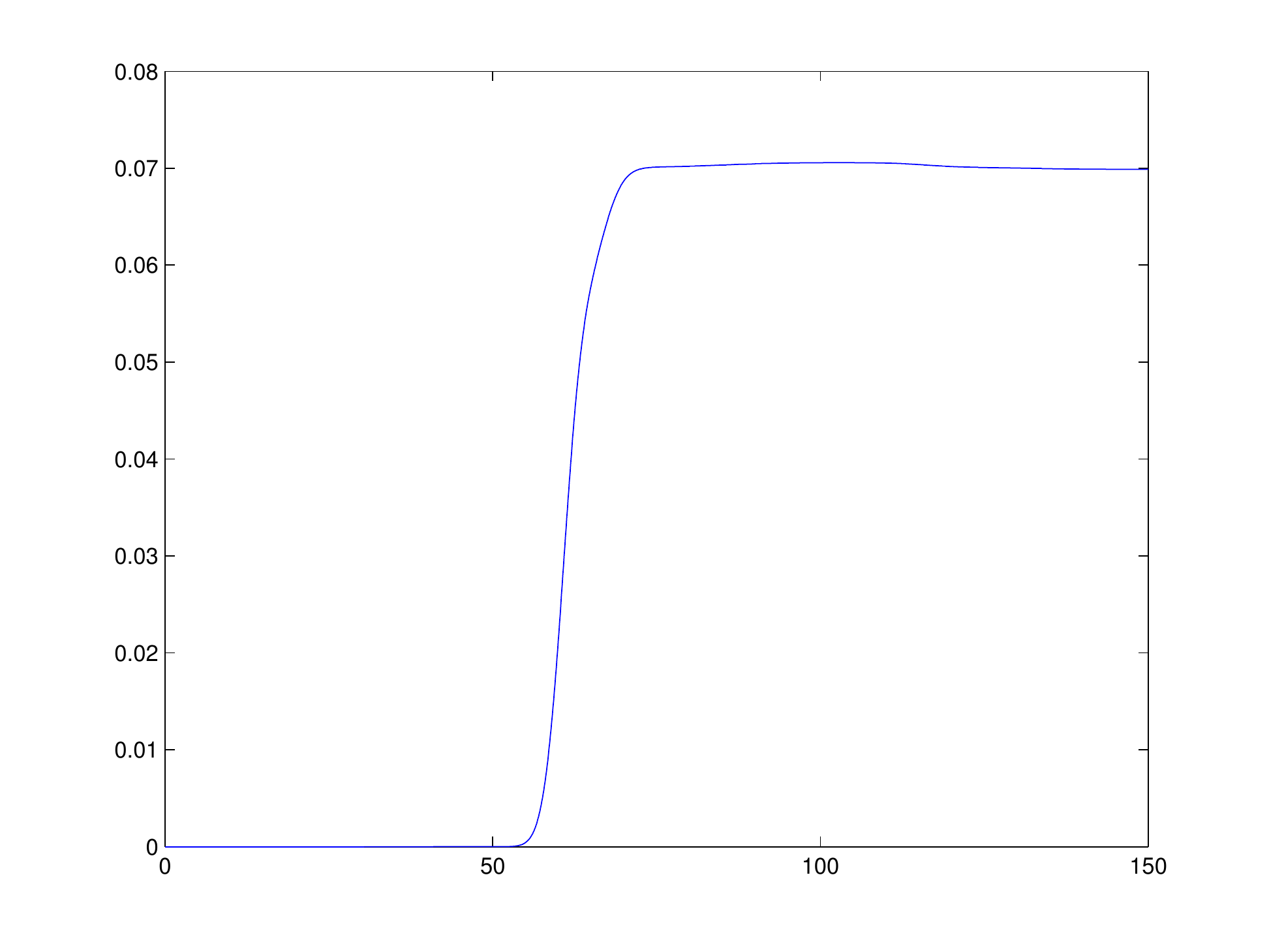} &
   \hspace{-0.4in}
   \includegraphics[width=.35\textwidth,height=7cm]{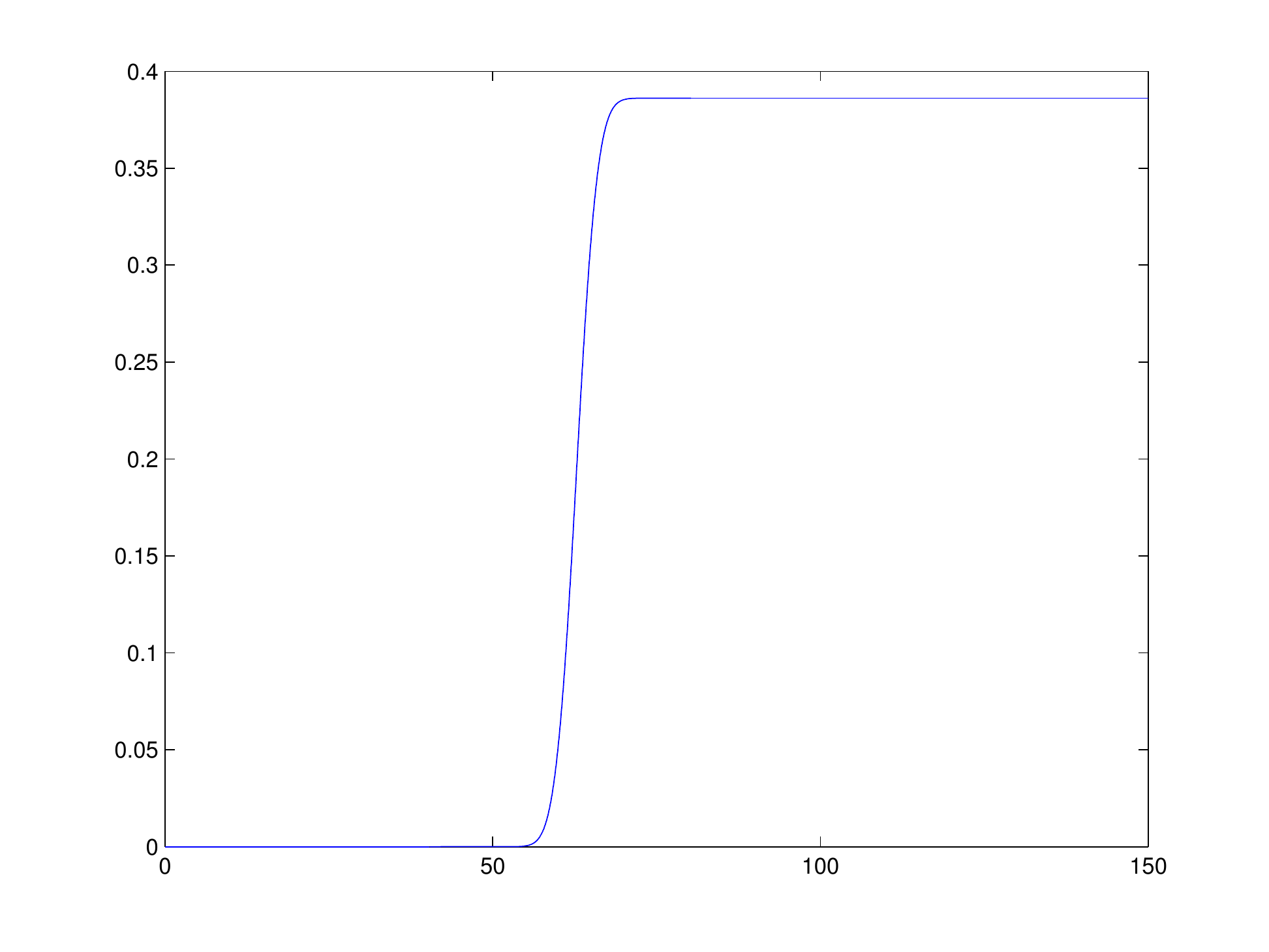} \\
   \hspace{-0.2in}
   \includegraphics[width=.35\textwidth,height=7cm]{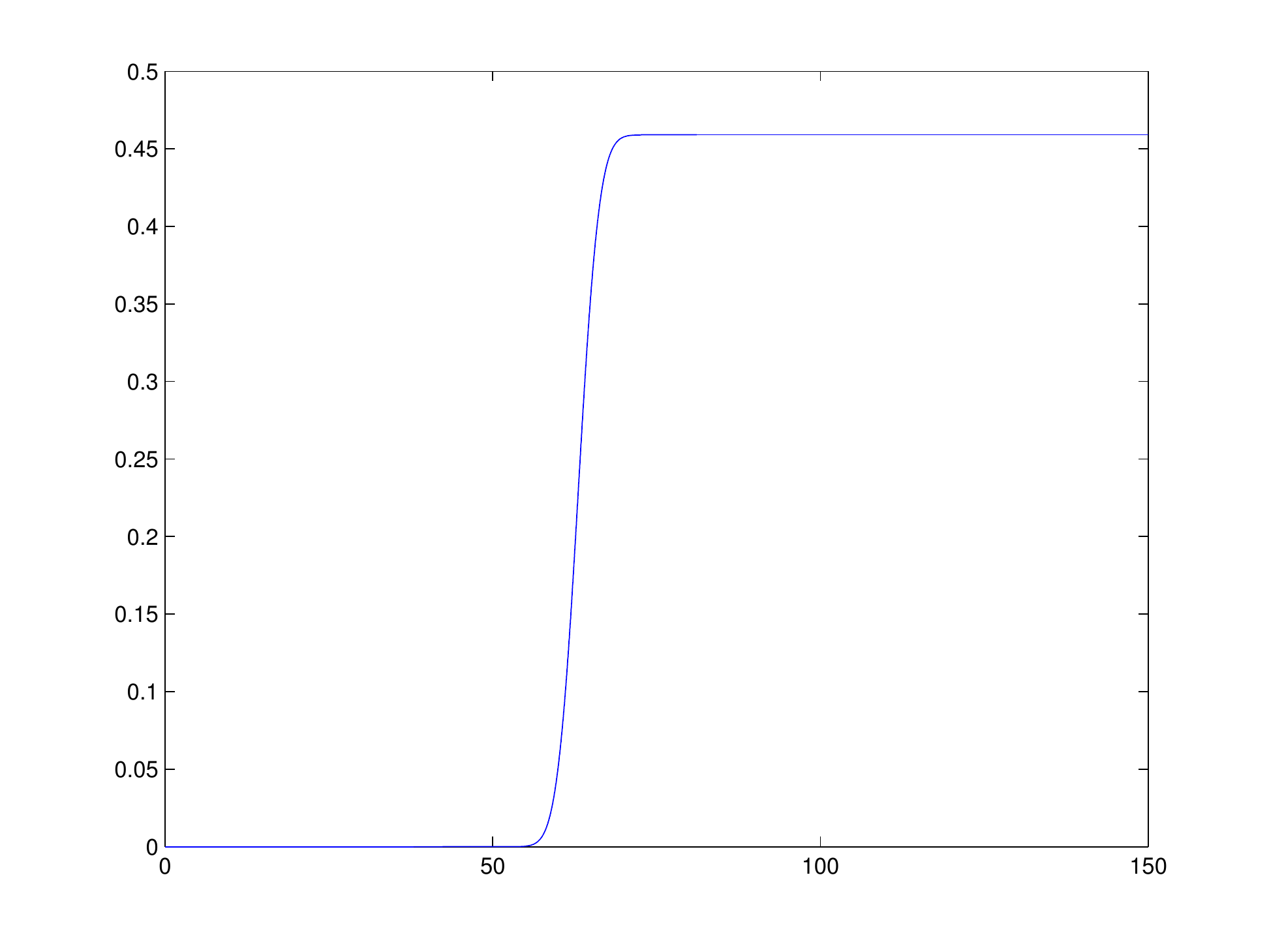} &
   \hspace{-0.4in}
   \includegraphics[width=.35\textwidth,height=7cm]{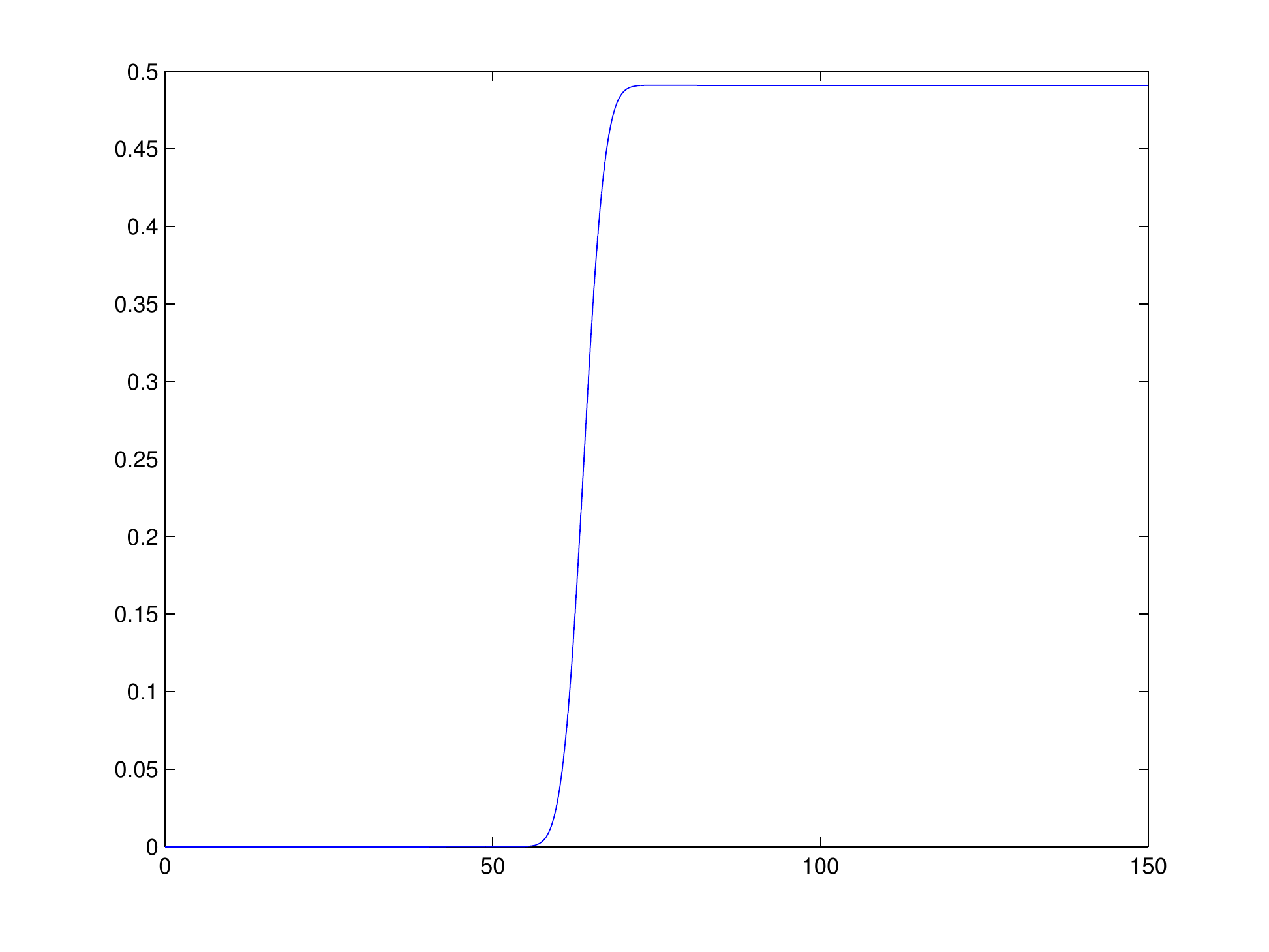}
  \end{tabular}
  \caption{Evolution of the energy gain for $\omega =0$, $5$, $20$, $50$ and $100$.} \label{RHNE2}
\end{figure}

We have already tested the behaviour at high energy of our numerical scheme with the incoming wave packets. We present here another test with oscillating Gaussian data within the effective ergosphere~:
\begin{eqnarray}
\phi_0 (r_*) &=& e^{i \omega r_*/\lambda} e^{-(r_*-r_*^0)^2/\lambda^2}  \, ,\\
\phi_1 (r_*) &=& 0 \, ,
\end{eqnarray}
with the same parameters as the flare above, i.e. on $[-50,50]$, with $r^0_* = -37.5$ and scaling factor $\lambda =5$, evolved up to time $t=150$. Such data do not have a dominant direction of propagation. We therefore expect that at high energy, the gain will stabilize at $0.5$.
For $\omega =0$, we obtain a gain larger than $1$. This has little meaning since the data do not belong to a subspace on which the energy is positive definite. Increasing the frequency leads to an ever faster stabilization of the gain towards values tending to $1/2$ and to a solution propagating along the outgoing and incoming radial null geodesics from the support of the data. The convergence of the asymptotic gain as the frequency increases is not as fast as for wave packets, probably due to the much smaller size of the support of the data. We show three examples of propagation of the solution corresponding to $\omega =0, 5, 10$, in Figure \ref{RHNE1} and five examples of the stabilization of the gain for $\omega = 0,5,20,50,100$ in Figure \ref{RHNE2}. The behaviour is in conformity with our expectations.

We have also checked that the final value ${\cal G}_\infty$ of the gain does not depend on the position at which we measure the outgoing flux (provided it is located to the right of the support of the data of course). This means that the energy current of the numerical solution is conserved, in other words, our numerical scheme preserves the energy.

\subsection{Conclusion}

The toy model that we have studied suggests that the driving force behind superradiance is the variation of the total potential. How this should be measured is not clear at present, although we can safely say that the $L^1$ norm of the derivative of the potential is inadequate, since we could have a sequence of potentials whose derivatives are smooth and tend in the sense of distributions towards a Dirac mass at the origin while having a bounded $L^1$ norm. When we smoothed out the step potential, we observed that hyperradiance disappeared but incoming wave packets still produced superradiance for a narrow enough smoothing.

We have successfully observed incoming wave packets with a superradiant behaviour outside a Reissner-Nordstr\o m black hole. This is because we have managed to choose physical parameters that produce a potential with a steep and rapid transition between its two limit values. Other numerical experiments that we have performed with different values of the physical parameters seem to indicate that superradiance becomes stronger as the black hole gets closer to the extreme case, i.e. as the ratio $M/\vert Q\vert$ approaches $1$ (while remaining larger than $1$). The maximum value of the gain ${\cal G}_\infty \simeq 1.449$ that we observe for the parameters \eqref{PhysicalParameters} is consistent with the findings of \cite{BriCaPa} regarding the reflection coefficient. One must however be careful in comparing our results with theirs. When investigating superradiance in the frequency domain, one studies a scattering process using a stationary approach. One looks for (infinite energy) solutions with a given frequency in time, satisfying outgoing radiation conditions at the horizon and either outgoing or incoming radiation conditions at infinity. The transmission and reflection coefficients $T(\omega)$ and $R(\omega)$ appear in the asymptotics of the solutions at the horizon and of the outgoing solution at infinity. This type of analysis reveals that superradiance occurs for $\omega < qQ/r_+$ and in \cite{BriCaPa}, the reflection coefficient is found to be always lower than $2$. In the time-domain approach, also called the time-dependent scattering approach, one deals with finite energy solutions propagating in from past null infinity and splitting into a part that falls into the black hole and another that propagates to future null infinity. The scattering matrix maps the scattering data in the past\footnote{For a complete scattering process, one would consider also data emerging from the black hole in the distant past. The scattering matrix would contain the two reflection coefficients and the two transmission coefficients associated with the incoming and the outgoing data. In the case of superradiance, we are only interested in the reflection coefficient for incoming fields from past null infinity, so we assume that nothing emerges from the black hole, we only have incoming data.} to the scattering data in the future. The reflection coefficient measures the ratio between the energy of the outgoing future scattering data and the incoming past scattering data. The bound obtained in \cite{BriCaPa} has a direct translation in the time-dependent picture as an upper bound on the energy gain~: ${\cal G}_\infty \leq 2$. In this paper, our study of the evolution of incoming wave packets is close to a scattering experiment and the gains we observe indeed remain lower than $2$. A feature of our observations which may appear surprising, in view of the frequency domain results, is that for the values of the physical parameters \eqref{PhysicalParameters}, for which $qQ/r_+ \simeq 0.97$, we find that the largest gain ${\cal G}_\infty \simeq 1.449$ is obtained for $\omega =2.3$. One must however remember that our $\omega$ describes the spatial frequency of our wave packet initial data and not the time frequency $\omega$ of the solutions in the stationary approach. Both frequencies have a range in which superradient behaviour is possible. The ranges are possibly related but it is not clear how. Besides, the range in which the spatial frequency allows superradiance depends on the profile of the wave packet and on the location of the data at $t=0$. It would be interesting to study the relations between the two frequencies and between the associated scattering matrices. To this day, it seems that such a study is missing in the literature. Also, even if we assumed that the ranges of time and spatial frequencies allowing superradiance were the same, we should remember that the spatial frequency of our wave packets is in fact $\omega/5$ due to our scaling parameter $\lambda =5$. Hence the superradient interval for our $\omega$ would be close to $[0,5]$, which is in fact not far from what we actually observe.

It would also be interesting to check the agreement of our numerical experiments with the low frequency regime ($M\omega << 1$) calculation of the transmission coefficient in \cite{RiSa}. This would require to take larger and larger values of the mass $M$ of the black hole and to check that the transmission coefficient tends to $-1$, or equivalently that the gain tends to $2$. This would also be in agreement with the large $qQ$ results of \cite{BriCaPa}.

We have also observed superradiance for flare-type initial data located within the effective ergosphere. The gains we obtain in this setting are much larger and strongly exceed the bound on the reflection coefficient predicted by the frequency analysis. There is in fact no contradiction. The situation we are considering with flare initial data is no longer a scattering experiment where the data in the past purely come from null infinity. A solution that, at a given time, vanishes identically and whose time derivative is compactly supported within the generalized ergosphere, is necessarily the superposition of a wave emerging from the black hole and another coming in from infinity. So the gain we observe between the energy of the data and that of the outgoing part is not just the reflection coefficient for incoming data, but also involves the transmission coefficient for outgoing data. Therefore the bound obtained in \cite{BriCaPa} does not apply here. It would be interesting to see whether similar phenomena do occur for rotation-induced superradiance.

In all our numerical tests, the energy gain stabilizes to a finite limit as time becomes large. Hyperradiance does not seem to occur in the Reissner-Nordstr\o m case~; fields and their local energy decay in time. This can be seen as numerical evidence of the existence of a complete scattering theory as was constructed in \cite{Ba}. This is also an indication of stability of the geometry at the linearized level.

\section{Acknowledgments}

The authors would like to thank the anonymous referees for very helpful comments and suggestions. JPN wishes to thank Alain Bachelot, Claudio Paganini and Lars Andersson for interesting and useful conversations and Claudio Paganini for pointing out our incorrect earlier use of the word ``dyadosphere''. This research was partly supported by ANR funding ANR-12-BS01-012-01.

\appendix

\section{Proof of lemma \ref{LemmaConsLawT}} \label{ProofLemma}

We first need to calculate the Christoffel symbols for the Reissner-Nordstr\o m metric. We perform the calculation for a general metric of the form
\[ g = F(r) \d t^2 - F(r)^{-1} \d r^2 - r^2 (\d \theta^2 + \sin^2 \theta \d \varphi^2 )\, .\]
The non zero Christoffel symbols of the Levi-Civita connection are
\begin{gather*}
\Gamma^0_{01} = \frac{F'}{2F} \, ,~ \Gamma^1_{00} = \frac{FF'}{2} \, ,~ \Gamma^1_{11} = -\frac{F'}{2F} \, ,\\
\Gamma^1_{22} = -rF \, ,~ \Gamma^1_{33} = -rF \sin^2 \theta \, , \\
\Gamma^2_{12} = \Gamma^3_{13} = \frac{1}{r} \, ,~ \Gamma^2_{33} = - \sin \theta \cos \theta \, ,~ \Gamma^3_{23} = \cot \theta \, .
\end{gather*}
We will also use the decomposition of the charged Klein-Gordon equation \eqref{CKGRNphi} in terms of the real and imaginary parts of the field $f$~:
\begin{eqnarray*}
F\square_g f_1 + 2 \frac{qQ}{r} \partial_t f_2 + ( Fm^2 - \frac{q^2Q^2}{r^2} ) f_1 =0 \, ,\\
F\square_g f_2 - 2 \frac{qQ}{r} \partial_t f_1 + ( Fm^2 - \frac{q^2Q^2}{r^2} ) f_2 =0 \, .
\end{eqnarray*}
The divergence of the stress-energy tensor is given by
\begin{eqnarray*}
\nabla^a T_{ab} &=& ( \square_g f_1 + (m^2 - q^2 A_a A^a ) f_1 ) \partial_b f_1 \hspace{0.5in} \mbox{(note that } A_a A^a = F^{-1} \frac{Q^2}{r^2} \mbox{)} \\
&& + ( \square_g f_2 + (m^2 - q^2 A_a A^a ) f_2 )\partial_b f_2 \\
&& - q^2 ( A^a \nabla_b A_a ) \vert f \vert^2 \\
&=& \frac{2qQ}{rF} ( \partial_t f_1 \partial_b f_2 - \partial_t f_2 \partial_b f_1 ) - q^2 ( A^a \nabla_b A_a ) \vert f \vert^2 \, .
\end{eqnarray*}
This does not vanish, but contracting this expression with $\partial_t$ (which is a Killing vector field), we obtain
\[ \nabla^a T_{a0} =\frac{2qQ}{rF} ( \partial_t f_1 \partial_t f_2 - \partial_t f_2 \partial_t f_1 ) - q^2 ( A^a \nabla_t A_a ) \vert f \vert^2 =0 \]
using the fact that
\[ \nabla_t A_a \d x^a = (\partial_t A_a - \Gamma^c_{0a} A_c ) \d x^a = (0 - \Gamma^0_{0a} A_0 ) \d x^a = - \Gamma^0_{01} \frac{Q}{r} \d r = -\frac{F'}{2F} \frac{Q}{r} \d r \]
and therefore $A^a \nabla_t A_a =0$. This concludes the proof of the lemma. \qed

\end{document}